\documentclass{statsoc}
\usepackage[section]{placeins}
\usepackage{amsfonts}
\usepackage{natbib}
\usepackage{graphicx}
\usepackage{subfigure}
\usepackage{psfrag}
\usepackage{newlfont}

\usepackage{pgfplots}

\newtheorem{theorem}{Theorem}[section]

\newtheorem{proposition}[theorem]{Proposition}

\title{Estimation and Testing  for Covariance-Spectral Spatial-Temporal Models}

\author[Ali M. Mosammam and John T. Kent]{Ali M. Mosammam; a.m.mosammam@znu.ac.ir}
\address{Department of Statistics, University of Zanjan, Zanjan, Iran.}
\email{a.m.mosammam@znu.ac.ir}
\author[Ali M. Mosammam and John T. Kent]{John T. Kent; j.t.kent@leeds.ac.uk}
\address{Department of Statistics,
University of Leeds, Leeds, LS2 9JT, UK.}
\email{j.t.kent@leeds.ac.uk}

\begin{document}

\maketitle

\begin{abstract}

  In this paper we explore a \emph{covariance-spectral modelling} strategy
  for spatial-temporal processes which involves a spectral approach
  for time but a covariance approach for space. It facilitates the
  analysis of coherence between the temporal frequency components at
  different spatial sites.  \cite{Stein2005} developed a
  semi-parametric model within this framework. The purpose of this
  paper is to give a deeper insight into the properties of his model
  and to develop simpler and more intuitive methods of estimation and
  testing.  An example is given using the Irish wind speed data.

\end{abstract}

\section{Introduction}\label{ch5sec1}
There is a need for tractable yet flexible spatial-temporal models
in applications such as environmental modelling.  Two natural
starting points are models for purely spatial or purely temporal
data. For example, one may consider time as an extra spatial
dimension; then spatial statistics techniques \citep{Cressie1993}
can be applied. However, this approach ignores the fundamental
differences between space and time such as coherence, which arises,
e.g.,  if a wind is blowing across a spatial region.  On the other
hand, starting from a time series perspective, one way to think of a
spatial-temporal process is as a multiple time series
\citep{Priestley1981} where the spatial locations of the data index
the components of the time series. However, this approach ignores
the regularity in space and does not allow inferences about the
process at sites where data are not observed.

In this paper we focus on a \emph{covariance-spectral} modelling strategy
which intertwines the roles of space and time in a deeper way.
Consider a real-valued stationary spatial-temporal process  $Z(s,
t)$ defined on
   $\mathbb{R}^d\times \mathbb{R}$  with covariance function $C(h,u)$,
where $\ h \in \mathbb{R}^d$ represents a spatial lag in $d$
dimensions, and $\ u \in \mathbb{R}$ represents a  temporal lag. The
covariance function has a spectral representation
\begin{eqnarray}\label{asf1574}
C(h, u)=FT_{ST}\{f(\omega,\tau)\}=\int e^{i(h'\omega+u\tau)}f(\omega , \tau)d\omega d\tau,
\end{eqnarray}
where for simplicity we usually assume the spectral measure has a
density $f(\omega,\tau)$ for $\omega \in \mathbb{R}^d, \ \tau
\in \mathbb{R}$. The subscripts ``$S$" and ``$T$" denote Fourier
transforms with respect to space and time respectively.  Taking a
``half Fourier transform" of $f$ over the spatial frequency yields
an intermediate function
\begin{eqnarray}\label{f1574}
H({h,\tau})=FT_{S}\{f(\omega,\tau)\}=\int e^{ih'\omega}f(\omega , \tau)d\omega,
\end{eqnarray}
so that
\begin{eqnarray*}
C(h, u)=FT_{T}\{H(h, \tau)\}=\int e^{iu\tau}H(h, \tau)d\tau.
\end{eqnarray*}
We shall call $H$ a ``covariance-spectral function'' since it
depends on the spatial lag $h$  and the temporal frequency $\tau$.
 Our modelling strategy will be to look for tractable and
flexible choices for $H$.

This paper is organized as follows. In Section \ref{S-T:Models},
general properties and special cases of $H$ are discussed, including
Stein's model.    An exploratory analysis of the Irish wind data is
carried out in Section \ref{ch5sec5}; this data set provides a test
case for the estimation and testing methods developed in Section
\ref{ch5sec6}.
\section{Stationary spatial-temporal models}\label{S-T:Models}
\subsection{General properties}
As described in the Introduction a stationary spatial-temporal
covariance structure can be represented equivalently in terms of a
covariance function $C(h,u)$, a spectral density $f(\omega,
\tau)$ or a covariance-spectral function $H(h,\tau)$.  In this
section we investigate the relationships between these
representations, and explore $H$ in more detail.

The following proposition based on standard Fourier analysis sets out
the properties possessed by each of these representations.

\begin{proposition}\label{cress}
  For an integrable real-valued function $f(\omega,\tau)$ on
  $\mathbb{R}^d\times \mathbb{R}$, let
  $C({h,u})=FT_{ST}\{f(\omega,\tau)\}$ and
  $H({h,\tau})=FT_{S}\{f(\omega,\tau)\}$.  Then the following are
  equivalent.
\begin{description}
  \item[(a)]  $C$ is an even ($C(h,u) = C(-h,-u)$),  real-valued
positive semi-definite (p.s.d) function.

\item[(b)] $f$ is an even ($f(\omega, \tau) = f(-\omega, -\tau)$)
  nonnegative function.

\item[(c)] $H(h, \tau)$ is an even ($H(h, \tau) = H(-h, -\tau)$)
complex-valued p.s.d. function and $H(h, \tau) = \bar{H}(-h,
\tau)$.
\end{description}
\end{proposition}

Statistical modelling strategies can be based on looking for
tractable choices in terms of either $C$, $f$ or $H$.  In this paper
we focus on $H$.  We need to find choices for $H$ satisfying
 (c) together with an integrability condition  $\int
H(0,\tau) d \tau < \infty$.

Recall that for two stationary time series, the coherence function
$\rho(\tau)$ gives the complex correlation in the spectral domain
between the two series at frequency $\tau$.  It is often convenient to
express $\rho(\tau) = |\rho(\tau)| \exp\{i {Arg} \rho(\tau)\}$ in terms
of the absolute coherence function $|\rho(\tau)|$ and the phase ${Arg}
\rho(\tau)$.  The phase of the coherence function determines the
extent to which one process leads or lags the other process.

In the spatial-temporal setting, the coherence function also depends
on the spatial lag $h$. It takes a convenient form in terms of the
covariance-spectral function,
\begin{eqnarray*}\label{fa156}
\rho(h, \tau)=\frac{H(h, \tau)}{\sqrt{H(0, \tau)H(0,\tau)}}=\frac{H(h, \tau)}{k(\tau)}.
\end{eqnarray*}
For a fixed spatial site $s$, $k(\tau)=H(0, \tau)$ is the spectral
density of the stationary time series $\{Z(s, t),\
t\in\mathbb{R}\}$. For any fixed spatial lag $h$, $\rho(h, \tau)$
is the coherence function of the two stationary time series $\{Z(s,
t),\ t\in\mathbb{R}\} $ and $\{Z(s+h, t),\ t\in\mathbb{R} \}$,
each with spectral density $k(\tau)$. For each $\tau\in \mathbb{R} $,
$h\neq 0$, we have $| \rho(h, \tau)|\leq1$ and $\rho(0,
\tau)=1$.

\subsection{Special Cases}
\emph{Separable models.}
If any of the following three equivalent conditions holds,
\begin{eqnarray*}
  C(h,u)=C_S(h)C_T(u),\,   f(\omega, \tau) = f_S(\omega) f_T(\tau),  \,   H(h, \tau) = C_S(h) f_T(\tau),
\end{eqnarray*}
then the covariance structure is said to be \emph{separable}.  In this
case the coherence function takes the form
\begin{eqnarray*}
\rho(h, \tau)=\rho_S(h) = C_S(h)/C_S(0),
\end{eqnarray*}
which is real-valued and does not depend on the frequency $\tau$.
Further $k(\tau) = C_S(0)f_T(\tau)$. Separability is a convenient
mathematical assumption but is usually far too stringent for
practical applications.

\emph{Fully symmetric models.} The assumption of full symmetry is less
restrictive than separability, but still imposes strong constraints on
the covariance structure.  The property of full symmetry can be
expressed in three equivalent ways:
\begin{eqnarray*}
C(h,u) = C(-h,u),\, f(\omega, \tau) &= f(-\omega, \tau), \,H(h, \tau)=H(-h, \tau).
\end{eqnarray*}

Hence, full symmetry is equivalent to the condition that the
coherence function $\rho(h, \tau)$ is real-valued.
\cite{Cressie1999} and \cite{Gneiting2002} have chosen real-valued
$H(h, \tau)$ for the covariance-spectral representation of
  spatial-temporal covariance
 functions; hence they
have obtained  fully symmetric covariance
 functions. \cite{rao2014frequency} investigate the use of the half-spectral
representation to carry out approximate maximum likelihood inference
under the assumption of full symmetry and isotropy.

\emph{Temporal frozen field models.} In some sense the opposite of
full symmetry is the frozen field model, in which a single time
series $Z_T(t)$ with covariance $C_T(u)$ and spectral density
$f_T(\tau)$ is observed at each spatial site, but subject to a
suitable temporal lag,
\begin{eqnarray*}
Z(s,t)= Z_T(t+v's)
\end{eqnarray*}
 where $v \in \mathbb{R}^d$ represents a spatial ``drift''. The
 covariance, generalized spectral density and covariance-spectral
 functions of the resulting process take the forms
\begin{eqnarray}\label{frozen}
C(h,u)&=C_T(u+v'h),\cr f(\omega, \tau)
&=f_T(\tau)\delta({\omega-v\tau}),\cr H(h,
\tau)&=f_T(\tau)e^{i v'h\tau},
\end{eqnarray}
where $\delta$ is the Dirac delta function, viewed here as a
generalized function of $\omega$ for each $\tau$. In this case the
coherence function $\rho(h, \tau)=e^{i v'h\tau}$ has absolute
value  one for all $h$ and $\tau$, reflecting the coherent
dependence of the two time series  $\{Z_T(t),\ t\in\mathbb{R}\} $
 and $\{Z_T(t+v's),\ t\in\mathbb{R}\} $ .
 There are two other types of frozen field model,
not considered here, which start with a
  spatial process $Z_S(s)$  or a  spatial-temporal process $Z_0(s,
  t)$, respectively; see e.g. \cite{Cox1988} and \cite{Ma2003a}.

\subsection{Stein's Covariance-Spectral Model}
The temporal frozen field model (\ref{frozen}) is too rigid in
practice, so we consider a more general model due to \cite{Stein2005}
\begin{eqnarray}\label{steincov}
H(h, \tau)=k(\tau)D(h\gamma(\tau))e^{i\theta(\tau)v'h},
\end{eqnarray}
where various components have the following interpretations.
\begin{itemize}
\item[(a)] The temporal spectral density $k(\tau)=k(-\tau)$ is a
  symmetric nonnegative integrable function on $\mathbb{R}$. It is
  identical to $f_T(\tau)$ in (\ref{frozen}).

\item [(b)] The latent spatial covariance function $D(h)=D(-h)$ is
  a real-valued positive definite function on $\mathbb{R}^d$ with
  $D(0)=1$.  Note that the coherence function of (\ref{steincov}) is
\begin{eqnarray}\label{steincoh}
\rho(h, \tau)=D(h\gamma(\tau))e^{i\theta(\tau)v'h},
\end{eqnarray}
with absolute value $|\rho(h, \tau)|=D(h\gamma(\tau))\leq 1$ for
$h\neq 0$. Thus $D$ governs how the absolute coherence decays with
increasing spatial lag up to a factor depending on temporal frequency.
For practical work we follow \cite{Stein2005} and assume $D$ takes the
specific form
\begin{eqnarray}\label{parametriccoh}
D(h)=e^{-|h|^p},\ 0<p\leq 2,
\end{eqnarray}
where $p$ is generally unknown, in order to simplify the estimation of
the remaining parts of the model.

\item [(c)] The temporal decay rate function
  $\gamma(\tau)=\gamma(-\tau)$ is a positive even function of $\tau
  \in \mathbb{R}$. It governs how the rate of decay of absolute
  coherence in $h$ depends on the temporal frequency $\tau$.  The
  simplest choice is the constant function $\gamma(\tau) = $ const.

\item [(d)] The temporal phase rate function $\theta(\tau)=-\theta(-\tau)$ is
  an odd function on $\mathbb{R}$. It governs how phase of the
  coherence function, ${Arg} \rho(h,\tau) = \theta(\tau)v'h$
  depends on temporal frequency. The simplest choice is the linear
  function $\theta(\tau) = b \tau$.

\item [(e)] Finally the unit vector $v$ specifies a direction for
  the spatial-temporal asymmetry.
\end{itemize}

The corresponding spectral density  of (\ref{steincov}) is
\begin{eqnarray}\label{asymmetricsdf}
f(\omega, \tau)=\frac{k(\tau)}{\gamma(\tau)}f_S\left(\frac{
\omega-v\theta(\tau)}{\gamma(\tau)}\right),
\end{eqnarray}
where $f_S(\omega)$ is the purely spatial spectral density of the
spatial covariance function $D(h)$. The vector $v\theta(\tau)$ in
the phase shift of the coherence function (\ref{steincov}) appears as
the location shift of the spectral density (\ref{asymmetricsdf}) and
the scaling function $\gamma(\tau)$ appears as a scale shift.

One way to motivate a simple special case of Stein's model is
through a combination of a separable model and a temporally  frozen
model.  If $D(h)=C_S(h)$, $\gamma(\tau)=1$, $k(\tau)=f_T(\tau)$
and $\theta(\tau)=\tau$ in (\ref{steincov}), then
\begin{eqnarray}\label{mycoherence}
C(h, u)&=C_S(h)C_T(u+v'h),\cr f(\omega, \tau)
&=f_S(\omega-v\tau)f_T(\tau),\cr
H(h,\tau)&=f_T(\tau)C_S(h)e^{i\tau v'h}.
\end{eqnarray}
However, Stein's approach  in (\ref{steincov}) allows a greater
degree of flexibility by allowing more choices for $\gamma(\tau)$
and $\theta(\tau)$.

Many statistical tests for separability have been proposed recently
based on parametric models, likelihood ratio tests and spectral
methods, e.g. \cite{Fuentes2006} and \cite{Mitchell2006}. In the
purely spatial context, \citet{Scaccia2005} and \citet{Lu2002}
developed tests for axial symmetry and diagonal symmetry.  These tests
are valid only under a full symmetry assumption.  A lack of full
symmetry in Stein's model can be carried out by examining whether
$\theta(\tau)=0$.  If $\theta(\tau)=0$, then the resulting
spatial-temporal covariance function is fully symmetric.  Furthermore,
if $\theta(\tau)=0$ and $\gamma(\tau)={const.}$, then the model is separable.

\section{Inference for Stein's covariance-spectral model}\label{ch5sec6}
In this section we investigate methods of inference for Stein's
covariance-spectral model (\ref{steincov}). The parameters are $p,\ 0<p\leq
2$ in the latent spatial covariance function (\ref{steincoh}) and
three functional parameters $k(\tau)$, $\gamma(\tau)$ and
$\theta(\tau)$. The goals are to develop methods for parameter
estimation, goodness of fit assessment and interpretation.

The estimation procedure has several steps, so it is helpful to set out
the general strategy and notation before the details are given.
\begin{itemize}
\item[(a)] Starting from the data, construct the empirical
  covariance-spectral function $\tilde{H}(h,\tau)$, as a  raw summary statistic
  of the data.

\item[(b)] Carry out preliminary nonparametric smoothing of
  $\tilde{H}(h,\tau)$ over $\tau$ to get a smoothed empirical
  covariance-spectral function $\tilde{\tilde{H}}(h,\tau)$.

\item[(c)] Transform the $H$ function to be linear in the unknown parameters,
and use regression analysis (with the transformed $\tilde{\tilde{H}}(h,\tau)$ playing the role of the response
variable) to estimate the parameters.  This strategy
is used first to estimate $k(\tau)$ and second to estimate jointly $p$ and
$\gamma(\tau)$.  The procedure to estimate $\theta(\tau)$ follows the same
general principles, but involves a preliminary estimate of $v$ based on
maximizing a certain ratio of quadratic forms.

In other words the basic
 strategy is to estimate $k(\tau)$, $\gamma(\tau)$ and
$\theta(\tau)$ is to match the smoothed  empirical
covariance-spectral function $\tilde{\tilde{H}}(h,\tau)$ to its
theoretical value in (\ref{steincov}) using regression methods,
after transforming (\ref{steincov})  to linearize the dependence of
$k(\tau)$, $\gamma(\tau)$ and $\theta(\tau)$ in turn, on $\tau$. The
regression can be either parametric or non-parametric.
\cite{Stein2005} has suggested parametric forms based on
trigonometric polynomials for $k(\tau)$, $\gamma(\tau)$ and
$\theta(\tau)$.   He maximizes the likelihood numerically, a
computationally intensive procedure due to the need to invert
large matrices. Our method can be seen as simpler and more graphical,
enabling visual judgments to be made about the model. \cite{Stein2005} constructed various plots to assess the goodness of fit of the
model. We use similar plots  to  estimate the parameters through a regression analysis. Both parametric and nonparametric regression models are accommodated
by this methodology. The effect and importance of each parameter then can be seen directly in the appropriate plot.

\item[(d)] Finally, it is necessary to estimate standard errors of the
parameters.
\end{itemize}
\subsection{Initial data processing}

Suppose the data $\{Z(s_i,t),\ t=1, \ldots, T\}$ are given at
irregular spatial locations $s_i,\ i=1, \ldots, S$ and equally-spaced integer
times $t=1, \ldots, T$.  For notational convenience suppose the data
have already been centered to have mean 0.  The first step is to
construct the half-Fourier transform of the data
\begin{eqnarray*}
J(s_i,\tau)=\sum_{t=1}^T Z(s_i,t)e^{-2\pi it\tau},\,  \tau=1/T, \ldots,
[T/2]/T.
\end{eqnarray*}

To distinguish the empirical and smoothed versions of
various quantities, we use to denote the empirical and the initial
smoothed versions, respectively.

The sample covariance-spectral function is defined to be
\begin{eqnarray*}
\tilde{H}(h,\tau)&=&\frac{1}{T}J(s_i,\tau)\bar{J}(s_j,\tau),\
h=s_i-s_j\neq 0,\\
\tilde{H}(0,\tau)&=&\frac{1}{S}\sum_{i=1}^S\tilde{H}_i(0,\tau),
\end{eqnarray*}
where $\bar{J}$ is the complex conjugate of $J$ and
$\tilde{H}_i(0,\tau)=\frac{1}{T}|J(s_i,\tau)|^2$.
Here $h$ ranges through the set of spatial lags $s_i-s_j$,
assumed for simplicity to have no replication except for
$h=0$. The sample temporal spectral density at site $s_i$ is
defined by $\tilde{k}_i(\tau)=\tilde{H}_i(0,\tau)$ and the sample
overall temporal spectral density by
\begin{eqnarray}\label{tildek}
\tilde{k}(\tau)=\tilde{H}(0,\tau).
\end{eqnarray}
  Define the sample coherence
and the sample phase for the process at two sites separated by spatial
lag $h=s_i-s_j\neq0$ by
\begin{eqnarray}
\tilde{\rho}(h, \tau) &=&\tilde{H}(h,
\tau)/\sqrt{\tilde{H}_i(0,\tau)\tilde{H}_j(0,\tau)}, \, h=s_i-s_j\neq 0\\
\tilde{D}(h, \tau)&=&|\tilde{\rho}(h, \tau)|, \hspace{.5cm} \tilde{g}(h, \tau) ={Arg}(\tilde{\rho}(h,
\tau)),\label{tilded}
\end{eqnarray}
so $\tilde{\rho}(h,\tau)=\tilde{D}(h, \tau)\tilde{g}(h, \tau)$.

\subsection{Initial smoothing}

Note that $\tilde{D}(h,\tau) = |\tilde{\rho}(h,\tau)| = 1$ for all
$h =s_i - s_j \neq 0$, making it useless as it stands for the
estimation of $D(h,\tau)$. To fix this problem, we propose that some
initial smoothing of $\tilde{H}(h,\tau)$ with respect to the time
frequency be carried out.  Denote the resulting ``initially smoothed''
sample covariance-spectral function by $\tilde{\tilde{H}}(h,\tau)$.
Similarly, the corresponding initially smoothed empirical temporal spectral
density $\tilde{\tilde{k}}(\tau)$, absolute phase
$\tilde{\tilde{D}}(h,\tau)$ and argument of phase
$\tilde{\tilde{g}}(h,\tau)$ are obtained by using
$\tilde{\tilde{H}}(h,\tau)$ instead of $\tilde{H}(h,\tau)$ in
equations (\ref{tildek})--(\ref{tilded}).

The exact amount of initial smoothing is not critical here.  Enough smoothing
is needed to make $\tilde{\tilde{D}}(h,\tau)$ suitable for estimation
purposes.  At the same time we do not want to mask any broad patterns
in the data which will be fitted later using parametric or
nonparametric models. It is also possible to include
tapering over time but for simplicity we have not done so here.

\subsection{Estimation of $k(\tau)$}
The initially smoothed temporal spectral density
$\tilde{\tilde{k}}(\tau)$ is a crude estimate of $k(\tau)$.  This estimate
can be refined in two ways, depending on whether we carry out
nonparametric or parametric modelling.
\begin{itemize}
      \item[(i)] (nonparametric) A simple way to estimate the function $k(\tau)$ is  simply
    by regressing $\tilde{\tilde{k}}(\tau)$
 on $\tau$ nonparametrically, e.g. using the Nadaraya-Watson  \citep{Nadaraya1964,Watson1964} estimator.
 A convenient
implementation in R is given by the functions tt{dpill} and tt{locpoly} in the package
tt{KernSmooth} \citep{Ruppert1995}.  Let  $\hat{k}_{{np}}(\tau)$ define the fitted
nonparametric estimate.


    \item[(ii)] (parametric)
\cite{Stein2005} suggested a parametric  fractional exponential
model
\begin{eqnarray}\label{ktaumodel}
\log k(\tau)=-\beta\log\sin(|\pi\tau|)+ \sum_{k=0}^{K_1}c_k\cos(2\pi
k\tau), \hspace{.5cm}\,  \tau=1/T, \ldots, [T/2]/T,
\end{eqnarray}
where the condition $0\leq\beta<1$ guarantees  the integrability of
$k(\tau)$ and allows for long-range dependence \citep[Ch.
6]{Bloomfield1973, Beran1994}. Here $K_1$ is assumed known for the
moment.  Assume  $\log\tilde{\tilde{k}}(\tau)$ equals the right hand
side (RHS) of (\ref{ktaumodel}) plus  independently and identically
distributed (i.i.d.)  errors. Then OLS regression yields estimates
$\hat{\beta}, \hat{c}_0, \ldots, \hat{c}_{K_1}$, which define a
fitted spectral density $\hat{k}_{{par}}(\tau)$.
\end{itemize}

\subsection{Estimation of $p$ and $\gamma(\tau)$}
 Under assumptions (\ref{steincov}) and (\ref{parametriccoh}) the
following transformation linearizes the dependence of
$D(h,\tau)=|\rho(h,\tau)|$ on the parameters $p$ and
$\gamma(\tau)$:
\begin{eqnarray}\label{newgamma12}
 \log(-\log( D(h,\tau)))=p\log(|h|)+p\log \gamma(\tau).
\end{eqnarray}
In terms of the data, we shall treat $ \log(-\log (\tilde{\tilde{D}}(h,
\tau)))$ as the dependent variable in a regression on the righthand side of
(\ref{newgamma12}) with i.i.d. normal errors.
Estimation can take two forms, depending on whether we carry out
nonparametric or parametric modelling.

\begin{description}

\item[(i)] (nonparametric) We propose estimation in two stages.
    Initially treat (\ref{newgamma12}) as a parallel-lines regression
    model on $\log(|h|)$, with common slope $p$ and with intercepts
    $p \log \gamma(\tau)$ depending on $\tau$. Fit the parameters by
    OLS and denote the resulting estimate of $\gamma(\tau)$ by
    $\hat{\gamma}_{{init}}(\tau)$.  For the second stage, regress
    $\hat{\gamma}_{{init}}(\tau)$ on $\tau$ nonparametrically to
    get $\hat{\gamma}_{{np}}(\tau)$ and $\hat{D}_{{np}}(h,
    \tau)$.

\item[(ii)] (parametric)
Following \cite{Stein2005}, one way to model the even non-negative
function $\gamma(\tau)$ is with trigonometric polynomials,
\begin{eqnarray}\label{gammamodel}
\log\gamma(\tau)= \sum_{k=0}^{K_2}a_k\cos(2\pi k\tau),
\end{eqnarray}
where $K_2$ is a pre-specified number of terms.  Then
 the log-log transformation linearizes the
 dependence of $D(h,\tau)$ on the parameters $p$ and $a_0, \ldots, a_{K_2}$.
\begin{eqnarray}\label{newgamma1}
 \log(-\log (D(h,\tau)))=p\log(|h|)+p\sum_{k=0}^{K_2}a_k\cos(2\pi k\tau).
\end{eqnarray}
Here  (\ref{newgamma1}) depends linearly on $\log(|h|)$ and
$\cos(2\pi k\tau),\  k=0, \ldots K_2$, where the slope $p$ is the
same for all $\tau$.   Fitting this model
by OLS leads to estimates $\hat{p}$ and $\hat{a}_0, \ldots,
\hat{a}_{K_2}$ which define $\hat{\gamma}_{{par}}(\tau)$ and
$\hat{D}_{{par}}(h, \tau)$.
\end{description}

\subsection{Estimation of ``drift" direction $v$ and the
phase $\theta(\tau)$ }

In (\ref{steincov}), recall
\begin{eqnarray}\label{thethet}
g(h,\tau)={Arg}(H(h, \tau))=\theta(\tau)v'h.
\end{eqnarray}
We propose estimating the parameters by regressing the smoothed empirical
phase function $\tilde{\tilde{g}}(h, \tau)$ on $h$ and $\tau$
using ordinary least squares.  However, there are two complications:
$g(h,\tau)$ is an angle, not a number, and the regression is nonlinear.

First we deal with the angular problem; that is, $g(h, \tau)$ is an
angular variable for each $h$ and $\tau$, and hence defined only up
to an integer multiple of $2\pi$. But since $g(h, \tau)$ is a
continuous function of $\tau$ for each $h$, it can also be regarded as
a real-valued function initialized by $\tilde{g}(h, 0)=0$.  That is, for
each $h$ and $\tau$ an unambiguous choice for the winding number can
be found.  Let $g_R(h,\tau)$ denote this real-valued extension of
$g(h,\tau)$.  Similarly, provided the noise is not too large, the
empirical phase $\tilde{g}(h, \tau)$ and its smoothed version
$\tilde{\tilde{g}}(h, \tau)$ can be unambiguously unwound to give
real-valued extensions $\tilde{g}_R(h, \tau)$ and
$\tilde{\tilde{g}}_R(h, \tau)$.

Next we regress $\tilde{\tilde{g}}_R(h, \tau)$ on $\theta(\tau)v'h$
on $v$ and $\theta(\tau)$.
Since the regression is nonlinear we proceed in two stages.  The first stage
produces an estimate of $v$ and an initial estimate of $\theta(\tau)$.
The second stage produces a more refined estimate of $\theta(\tau)$.

Here are the details.  Ordinary least squares estimation involves
minimizing the sum of squares
\begin{eqnarray}\label{sse}
SSE=\sum_{\tau}\sum_{h\neq0}\left(\tilde{\tilde{g}}_R(h,
\tau)-\theta(\tau)v'h\right)^2.
\end{eqnarray}
 If $v$ is known,  then for each fixed $\tau$,
 the OLS estimate of
 $\theta(\tau)$ is given by
\begin{eqnarray}\label{ols}
  \hat{\theta}_{{init}}(\tau;v) &=& \frac{v'\sum_{h\neq0}\tilde{\tilde{g}}_R(h,
\tau)h}{v'Av},
\end{eqnarray}
where $A=\sum_{h\neq0}hh'$ is a $d\times d$ matrix.
Inserting (\ref{ols}) into (\ref{sse}) yields the reduced sum of
squares
\begin{eqnarray}\label{secondsse}
SSE(v)=\sum_{\tau}\sum_{h\neq0}\tilde{\tilde{g}}_R(h,
\tau)^2-\frac{v'Bv }{v'Av},
\end{eqnarray}
where $B=\sum_{\tau}\beta(\tau)\beta'(\tau)$ is a $d\times d$
matrix defined in term of the $d$-dimensional vector
$\beta(\tau)=\sum_{h}\tilde{\tilde{g}}_R(h, \tau)h$.

Minimizing (\ref{secondsse}) now reduces to an optimization problem
for a ratio of quadratic forms. The optimal $v$ is given by the
eigenvector corresponding to the largest eigenvalue of $A^{-1}B$
(e.g. \citet[p. 479]{Mardia1979}). Let $\hat{v}$ denote the result. Once $v$
has been estimated,  then the regression equation reduces to
\begin{eqnarray*}\label{thethet2}
g_R(h, \tau)=\theta(\tau)\hat{v}'h
\end{eqnarray*}
and the initial estimate of $\theta(\tau)$ becomes
\begin{eqnarray*}
  \hat{\theta}_{{init}}(\tau) &=& \frac{\hat{v}'\sum_{h\neq0}\tilde{\tilde{g}}_R(h,
\tau)h}{\hat{v}'A\hat{v}}.
\end{eqnarray*}

We can get a refined  estimate of
 $\theta(\tau)$ as follows.
\begin{description}
    \item[(i)] (nonparametric) Regress
$\hat{\theta}_{{init}}(\tau)$ on $\tau$ nonparametrically to
get $\hat{\theta}_{{np}}(\tau)$.
       \item[(ii)] (parametric) Regress $\hat{\theta}_{{init}}(\tau)$ on
$\tau$ parametrically to get $\hat{\theta}_{{par}}(\tau)$.
Following \cite{Stein2005}, one way to model  the  odd   function
$\theta(\tau)$ is with trigonometric polynomials,
\begin{eqnarray}\label{thetamodel}
\theta(\tau)= \sum_{k=1}^{K_3}b_k\sin(2\pi k\tau),
\end{eqnarray}
for some  fixed $K_3$.
\end{description}

\subsection{Estimation of Standard Errors}
In estimating both  functional parameters $\gamma(\tau)$ and
$\theta(\tau)$  we must do initial smoothing. In the first case  we
do initial smoothing  to ensure that the absolute coherence is less than one and
in the latter case to ensure the winding of the phase angle varies smoothly with $\tau$. Although there is no need for initial smoothing of
the empirical spectral density, we have initially smoothed  the
empirical spectral density to unify our estimation procedures.

But initial smoothing leads to autocorrelated errors, underestimated
standard errors and minor shift in the intercepts. Although ignoring
correlation usually introduces little bias in the estimates of
regression coefficients, it can introduce substantial bias in the
estimates of standard errors and this may lead to incorrect
inferences about OLS estimates. To resolve this problem, we assume an
approximate model for the autocorrelation of the initially smoothed
residuals.  The general procedure can be described as follows.  Consider  a general  linear
regression model $Y=X\beta+\varepsilon,
E(\varepsilon)=0, {cov}(Y)=\Sigma$, where  $Y$ is an
$SF\times 1$ random vector where $F=[T/2]$ and $\beta$ is a vector
of unknown regression parameters. Under the separability assumption
we have $\Sigma=\Sigma_S\otimes \Sigma_F$, where
   $\Sigma_S=(\sigma_{i,j})_{i,j=1}^S$ and $\Sigma_F=(\delta_{i,j})_{i,j=1}^F$ are covariance matrices
   and $\otimes$ denotes the Kronecker
   product. The OLS estimates are given by  $\hat{\beta}=(X'X)^{-1}X'Y$.

Under the separability  assumption we have
   ${cov}(Y_{i_1,j_1},Y_{i_2,j_2})=\sigma_{i_1,i_2}\delta_{j_2-j_1}$.
    Let
$\hat{\varepsilon}$ be the $SF\times 1$ vector of the
corresponding OLS residuals. The corresponding estimates are given
by
\begin{eqnarray*}
  \hat{\delta}_u&=&\frac{1}{SF}\sum_i\sum_{j=1}^{F-u}\hat{\varepsilon}_{i,j}\hat{\varepsilon}_{i,j+u}/
  \hat{\sigma}_{i,i},
  \cr
 \hat{\sigma}_{i_1,i_2}&=&\frac{1}{F }\sum_{j}\hat{\varepsilon}_{i_1,j}\hat{\varepsilon}_{i_2,j},
  \end{eqnarray*}
  where without loss of generality we  scale the covariance matrices
so that
  $\hat{\delta}_0=1$.
The variance of regression coefficients are given by
\begin{eqnarray*}
{var}(\hat{\beta})=(X'X)^{-1}X'(\hat{\Sigma}_S\otimes
\hat{\Sigma}_F)
   X(X'X)^{-1}.
\end{eqnarray*}
Note that for large values of $S$ and $F$ calculating the Kronecker
product $\hat{\Sigma}_S\otimes \hat{\Sigma}_F$ first and then
multiplying it by $X$ is   not only computationally inefficient, it
is also not feasible due to memory problems in $R$. To resolve this problem we use
the fact that $\hat{\Sigma}_S\otimes
\hat{\Sigma}_F=(\hat{\Sigma}_S\otimes I_F)(I_S\otimes
\hat{\Sigma}_F)$. These identity matrices make the calculation
easier because it is easy to see that sub-blocks of $X$  are
multiplied by $\hat{\Sigma}_F$ so we multiply each block separately. Here we also are approximating  $\hat{\Sigma}_F$  by a simple
Toeplitz matrix based on an AR(1) process. The covariance matrix for the stationary
AR(1) process has a well-known Toeplitz structure making computations simpler.
Furthermore calculating $(X'X)^{-1}$ can be also simplified because
of present of trigonometric functions in our regression models.

\subsection{General Considerations}
   \begin{itemize}
\item
When fitting trigonometric polynomials for functional parameters
$\kappa(\tau)$, $\gamma(\tau)$ and $\theta(\tau)$, it is necessary
to choose values for $K_1,K_2,K_3$. This choice can be made
subjectively or by some model selection criteria such as AIC or BIC.
In our example we will use AIC.

 \item We can construct pointwise  confidence intervals  for $\kappa(\tau)$,
$\gamma(\tau)$ and $\theta(\tau)$  using asymptotic properties of
the spectral density,  coherence and phase functions
\citep{Bloomfield1976}.

\end{itemize}

\section{Application to the Irish Wind Data}\label{ch5sec5}
The Irish wind data set is used here to provide a test case for the
estimation and testing methods developed in Section \ref{ch5sec6}.
The data  consist of average daily wind speeds (meters per second)
measured at 11 synoptic meteorological stations located in the
Republic  of Ireland during the period 1961-78, with 6,574
observations per location.
 Following \cite{Haslett1989} we  take a square root transformation to stabilize
the variance over both stations and time periods for each day of the
year and subtract the seasonal effect from the data.

\cite{Gneiting2002}, by plotting spatial-temporal correlations for
different spatial and temporal lags, has shown  that the wind speeds
measured at different stations are highly correlated and the
correlations decay substantially as  spatial or temporal lag
increases.  \cite{De2005}  by plotting the correlation function in
different directions concluded that the data have an isotropic
correlation structure.

\cite{Gneiting2002} used this data set to illustrate the lack of the
separability and full symmetry assumptions.  Indeed winds in Ireland
are predominantly westerly; hence  for different temporal lags the
west-to-east correlation of wind speed of
 stations will be  higher than the  east-to-west correlation; that is $C(h,u)>C(-h,u);\
 h'=(h_1,0),\ h_1>0, u>0$.
\begin{figure}
    \centering
\includegraphics[width=.325\columnwidth]{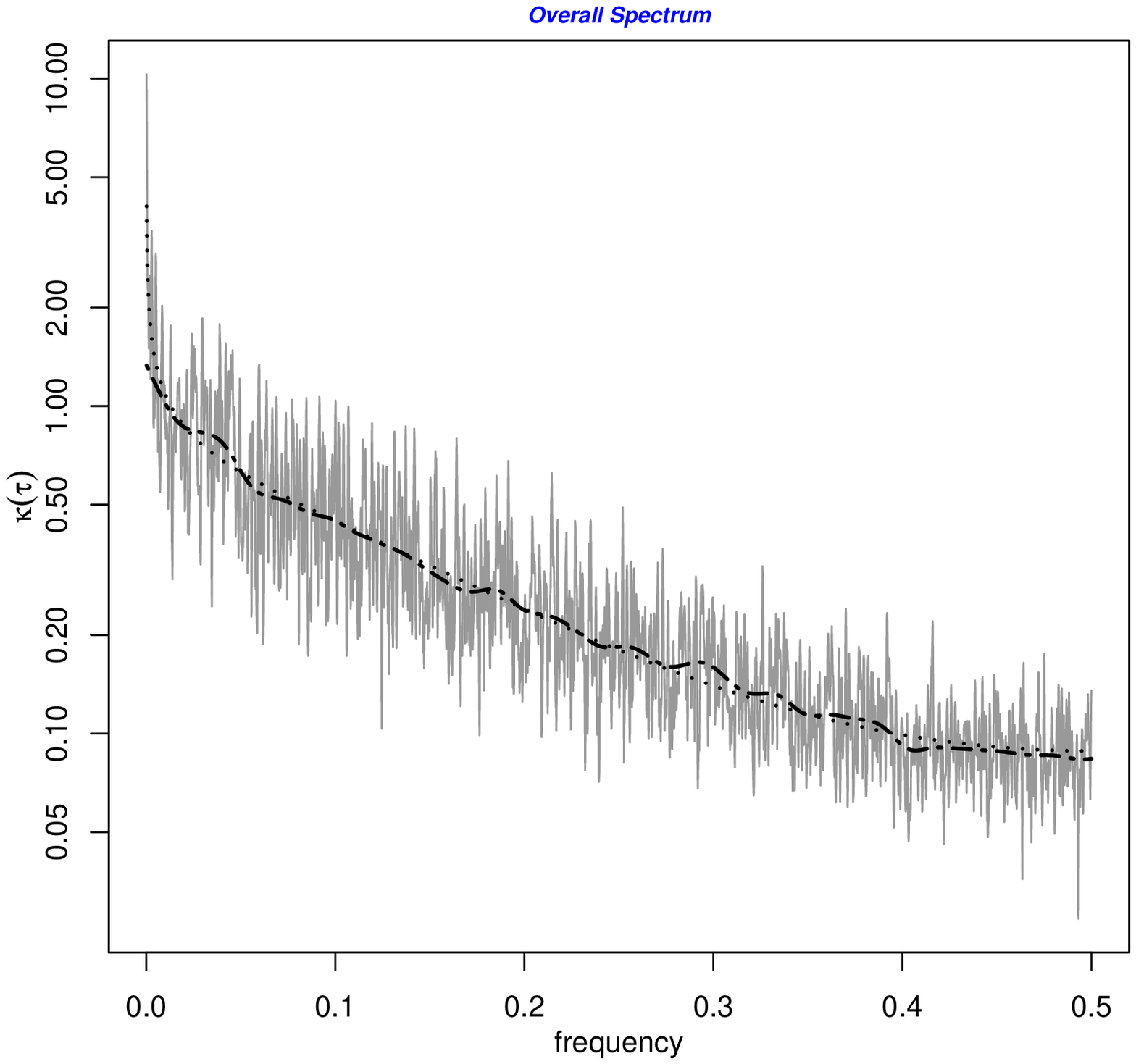}
\includegraphics[width=.325\columnwidth]{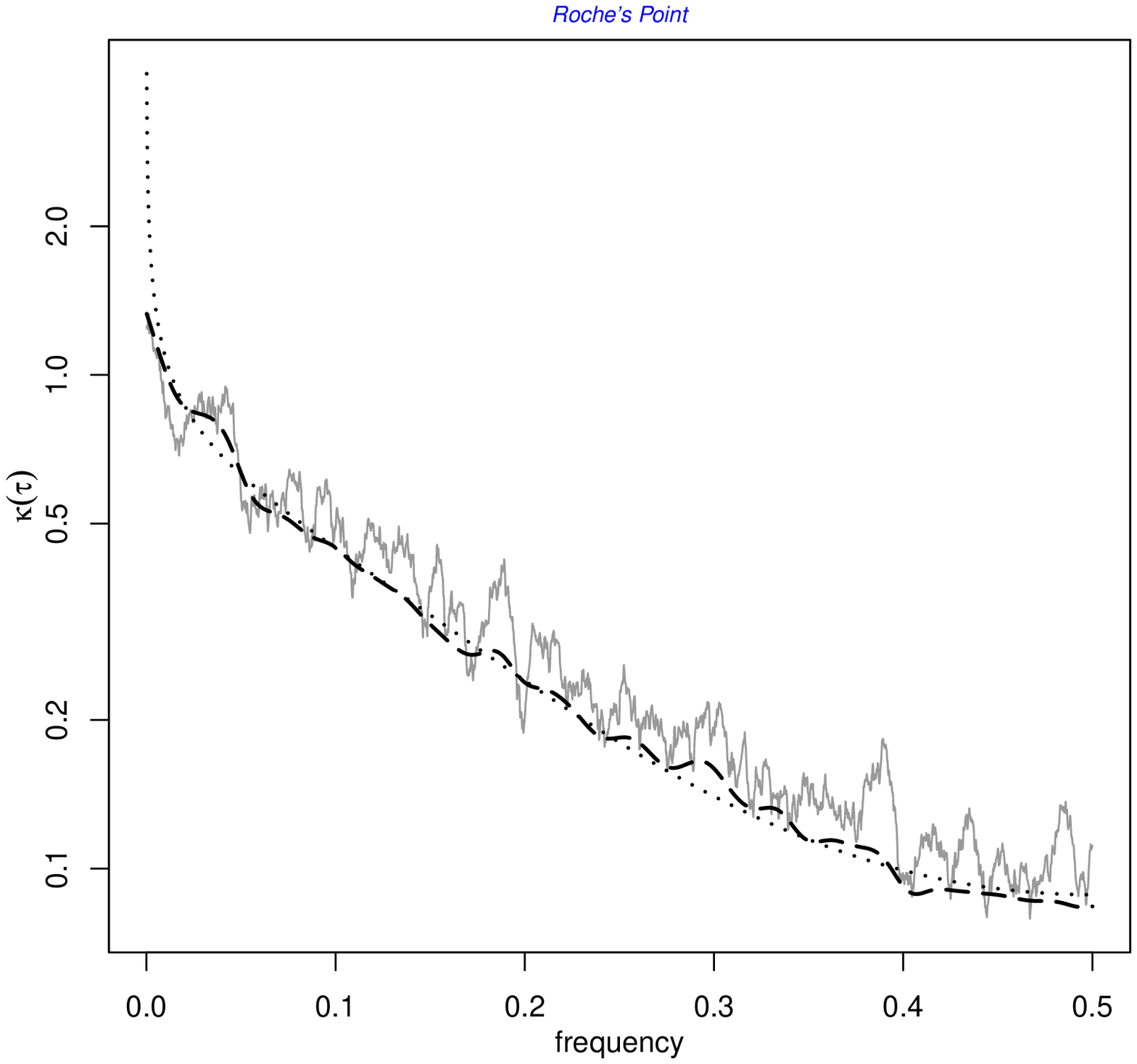}
\includegraphics[width=.325\columnwidth]{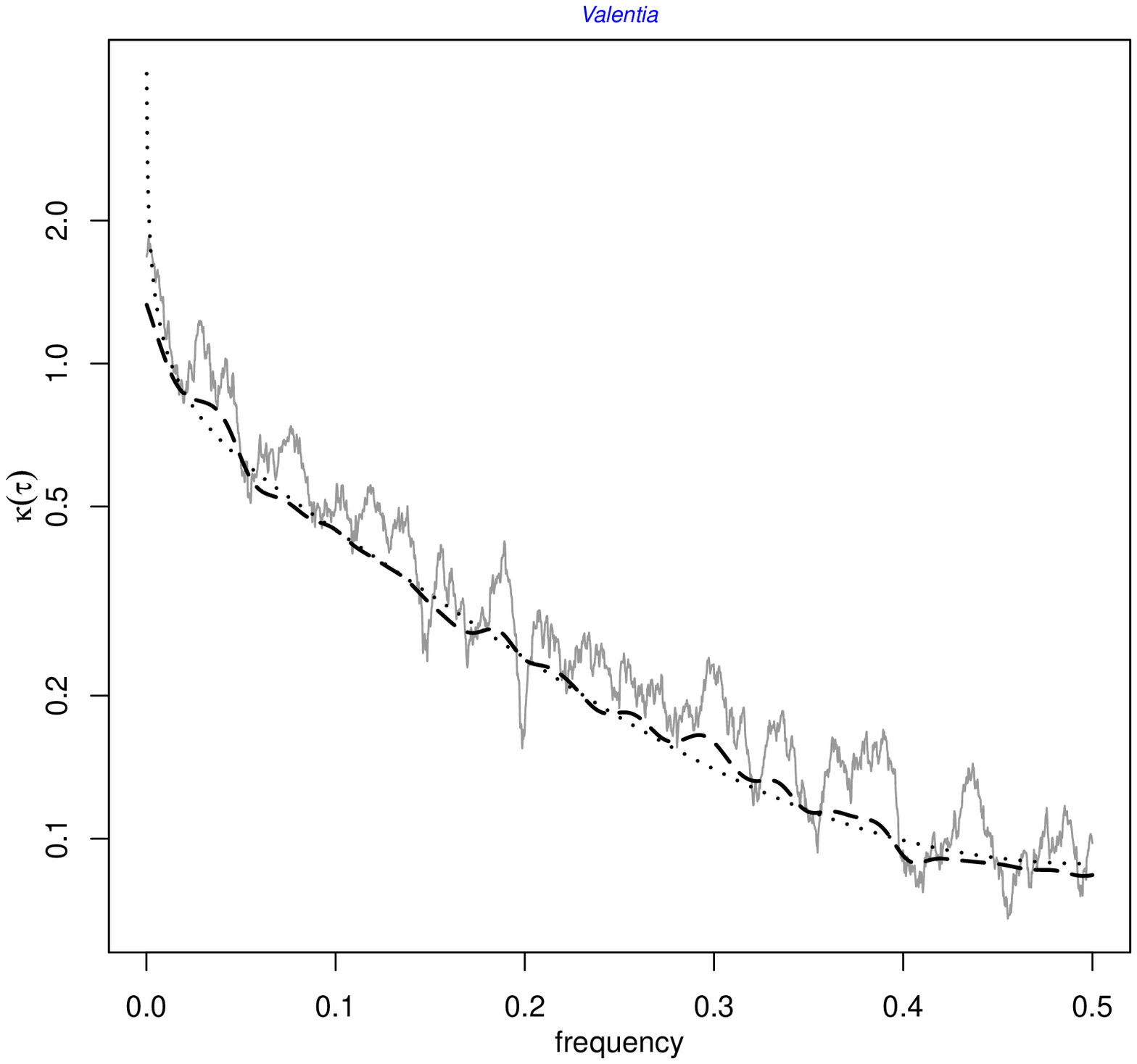}
\includegraphics[width=.325\columnwidth]{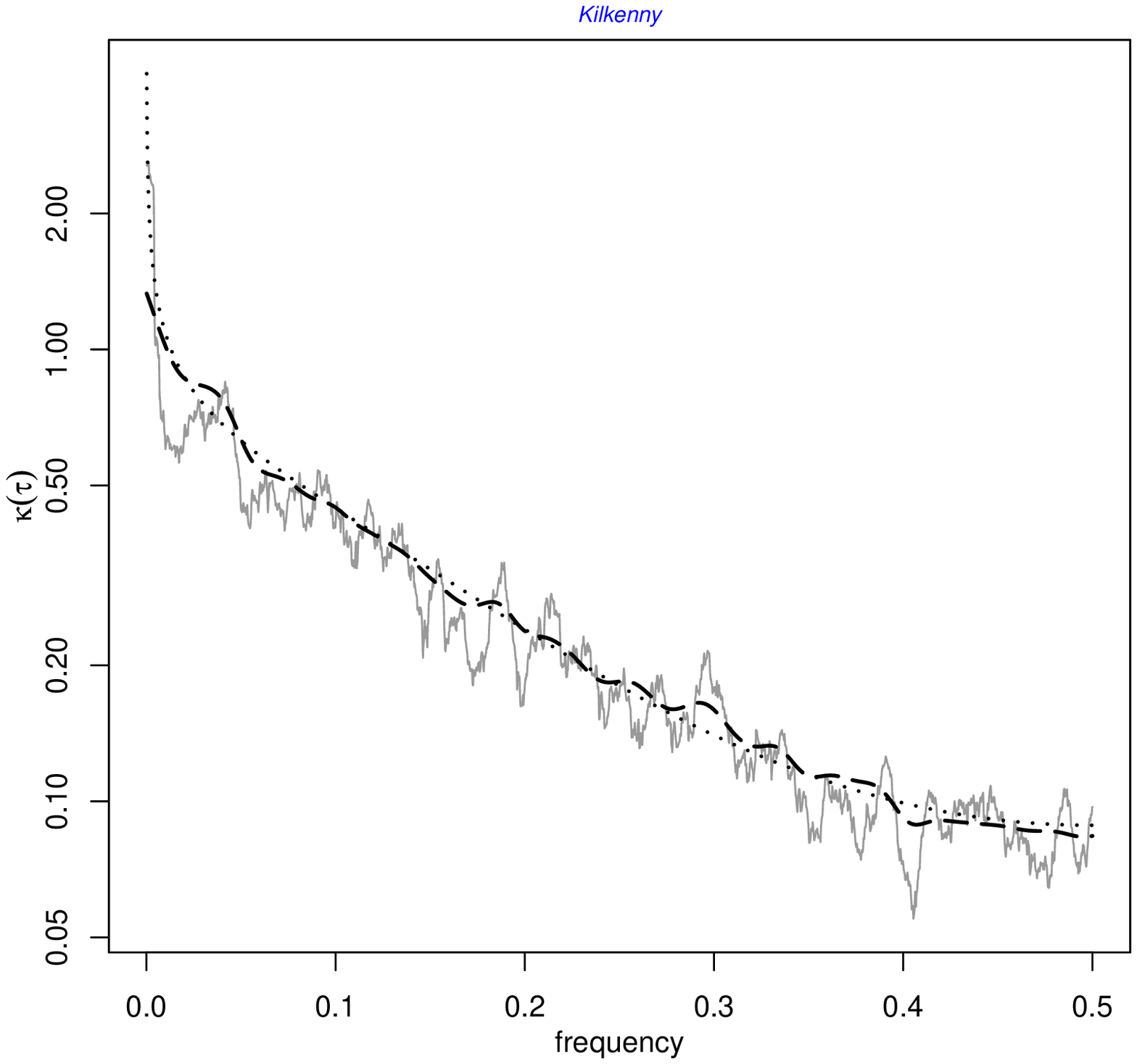}
\includegraphics[width=.325\columnwidth]{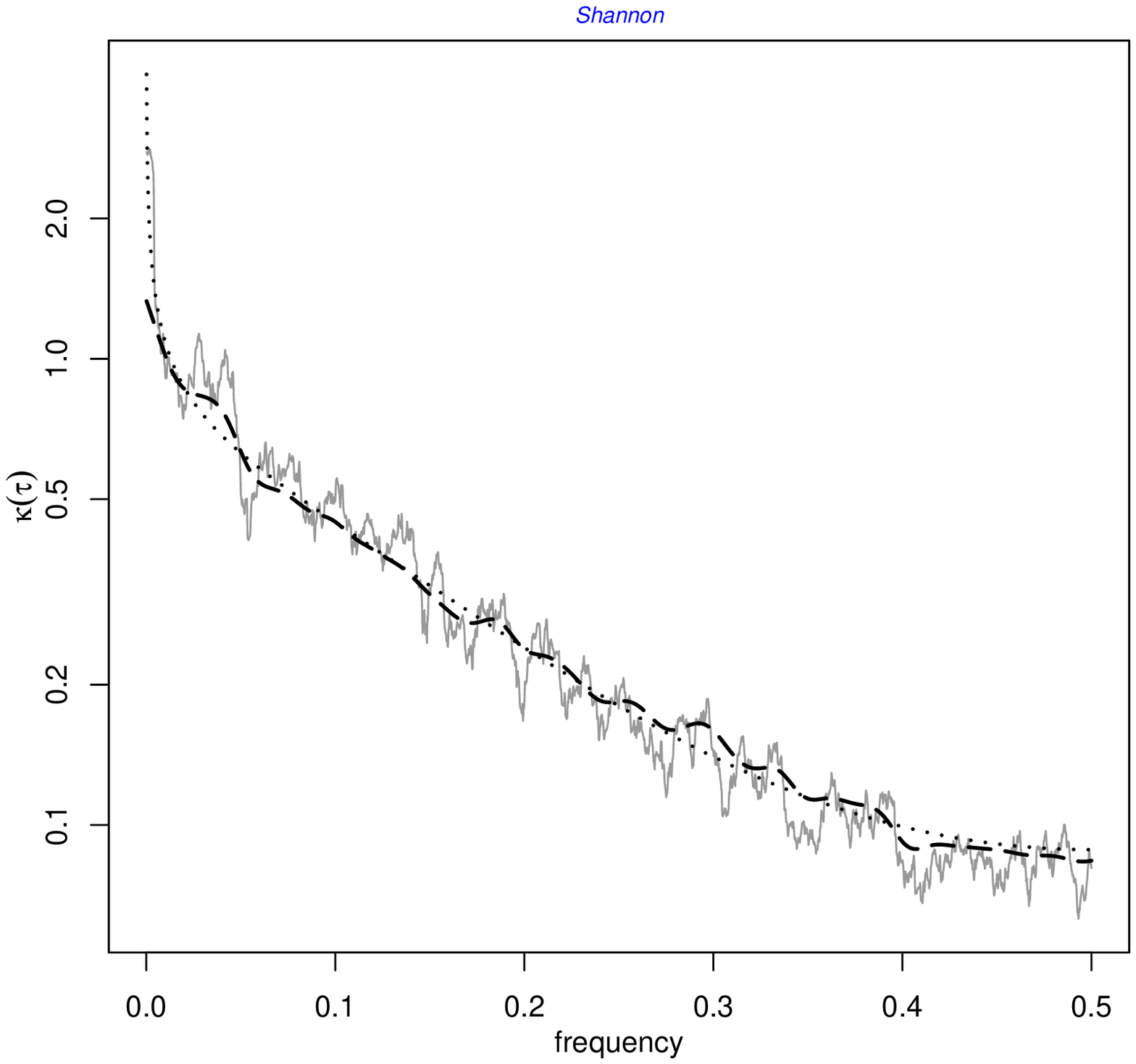}
\includegraphics[width=.325\columnwidth]{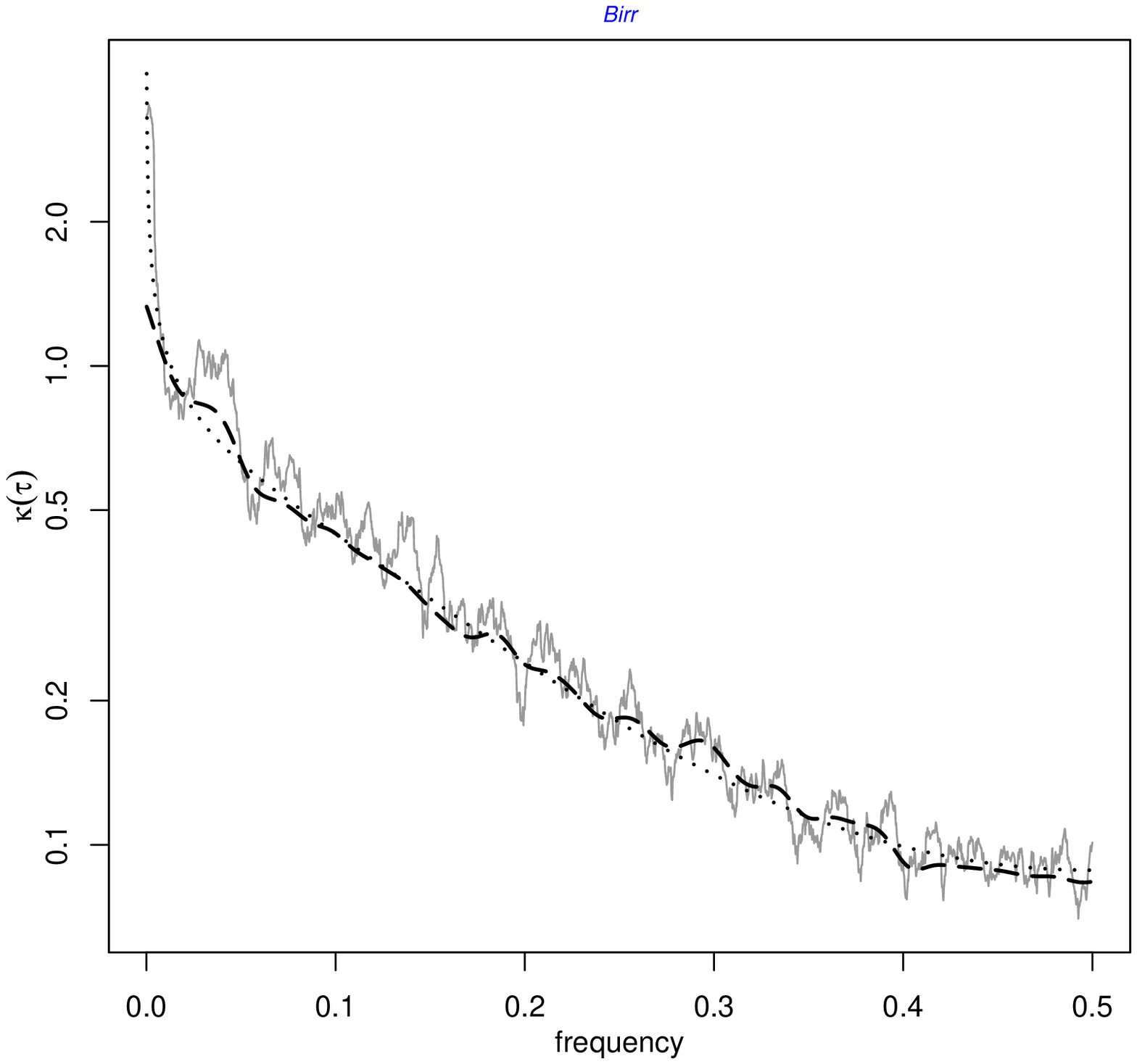}
\includegraphics[width=.325\columnwidth]{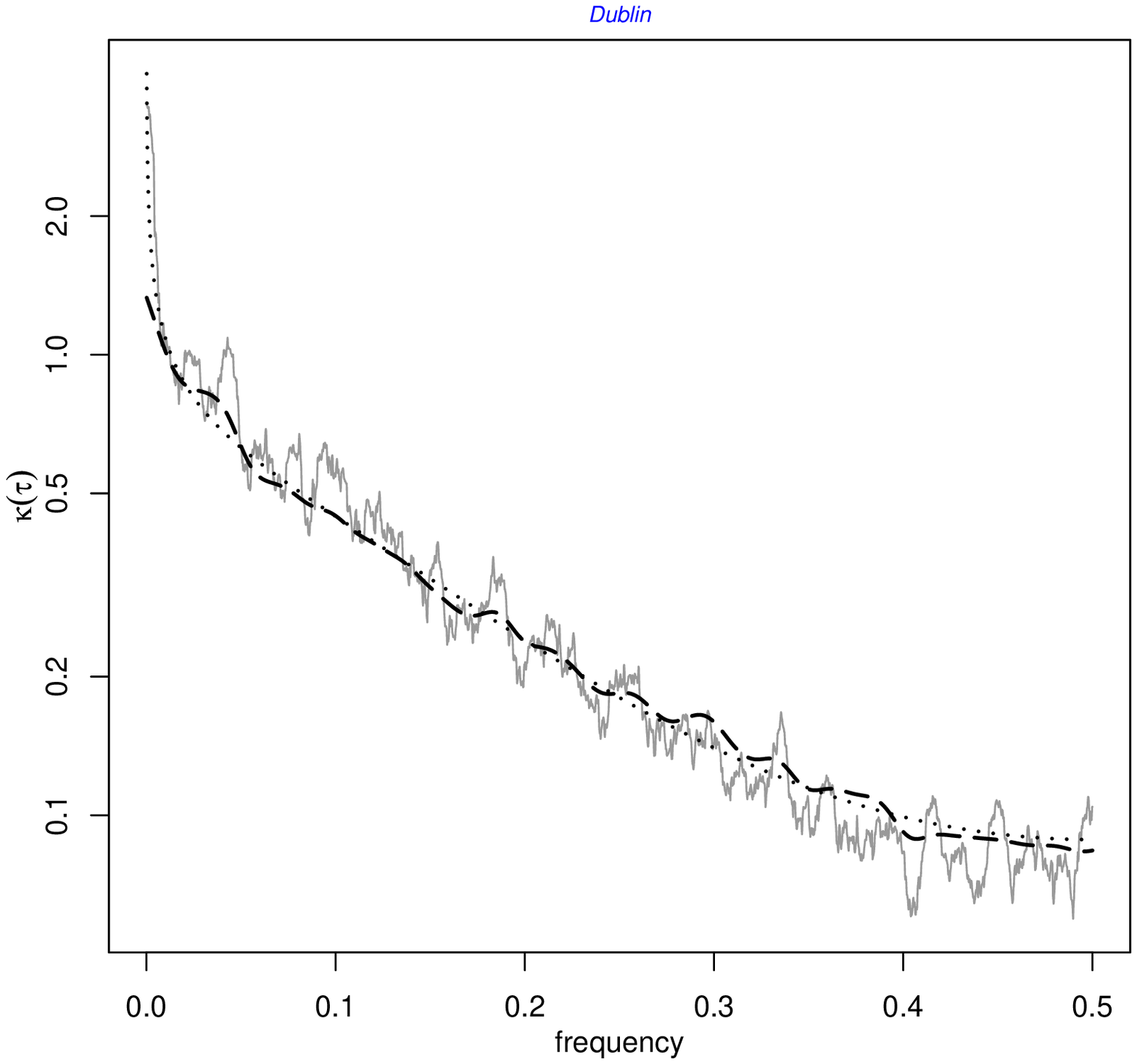}
\includegraphics[width=.325\columnwidth]{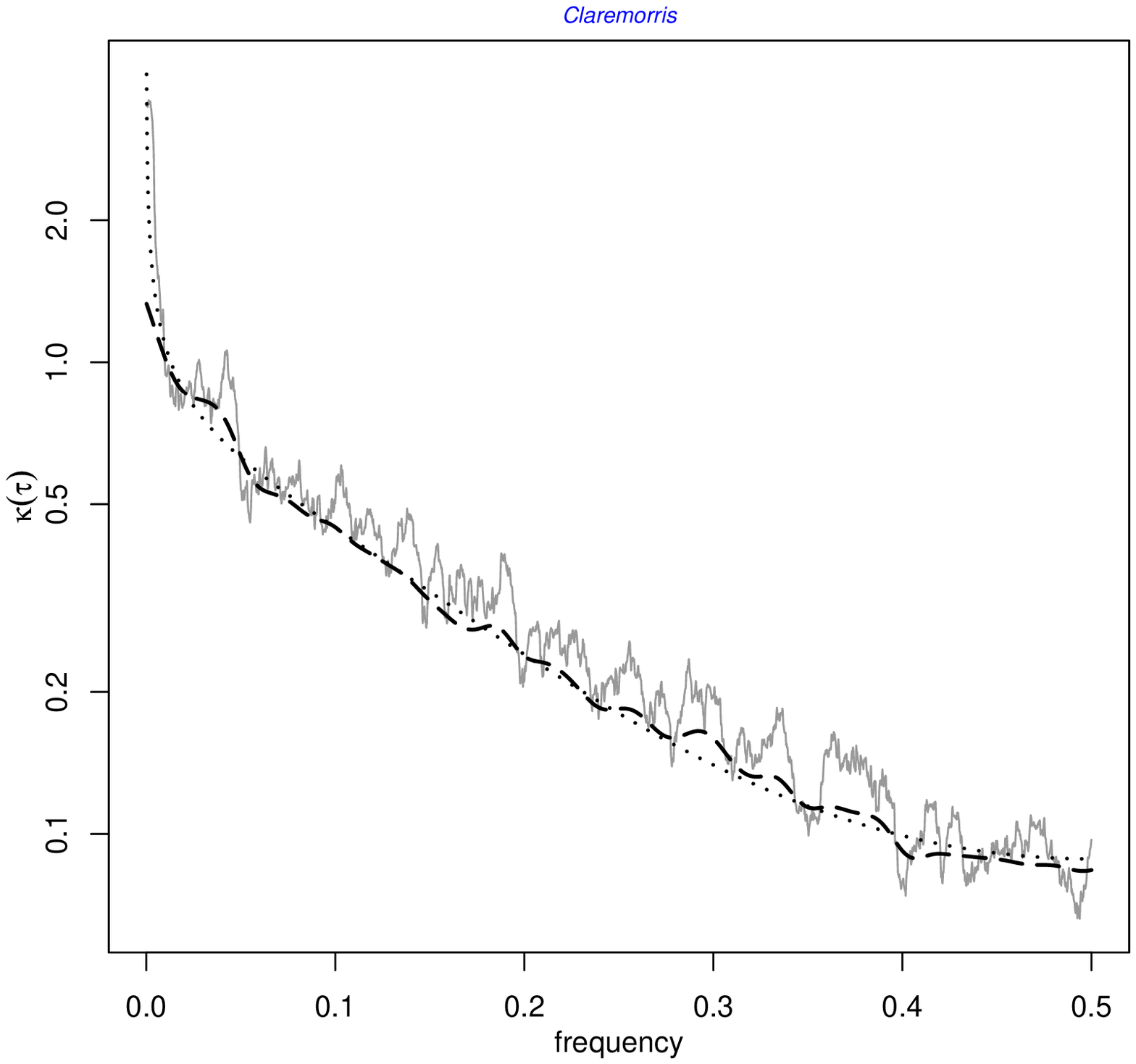}
\includegraphics[width=.325\columnwidth]{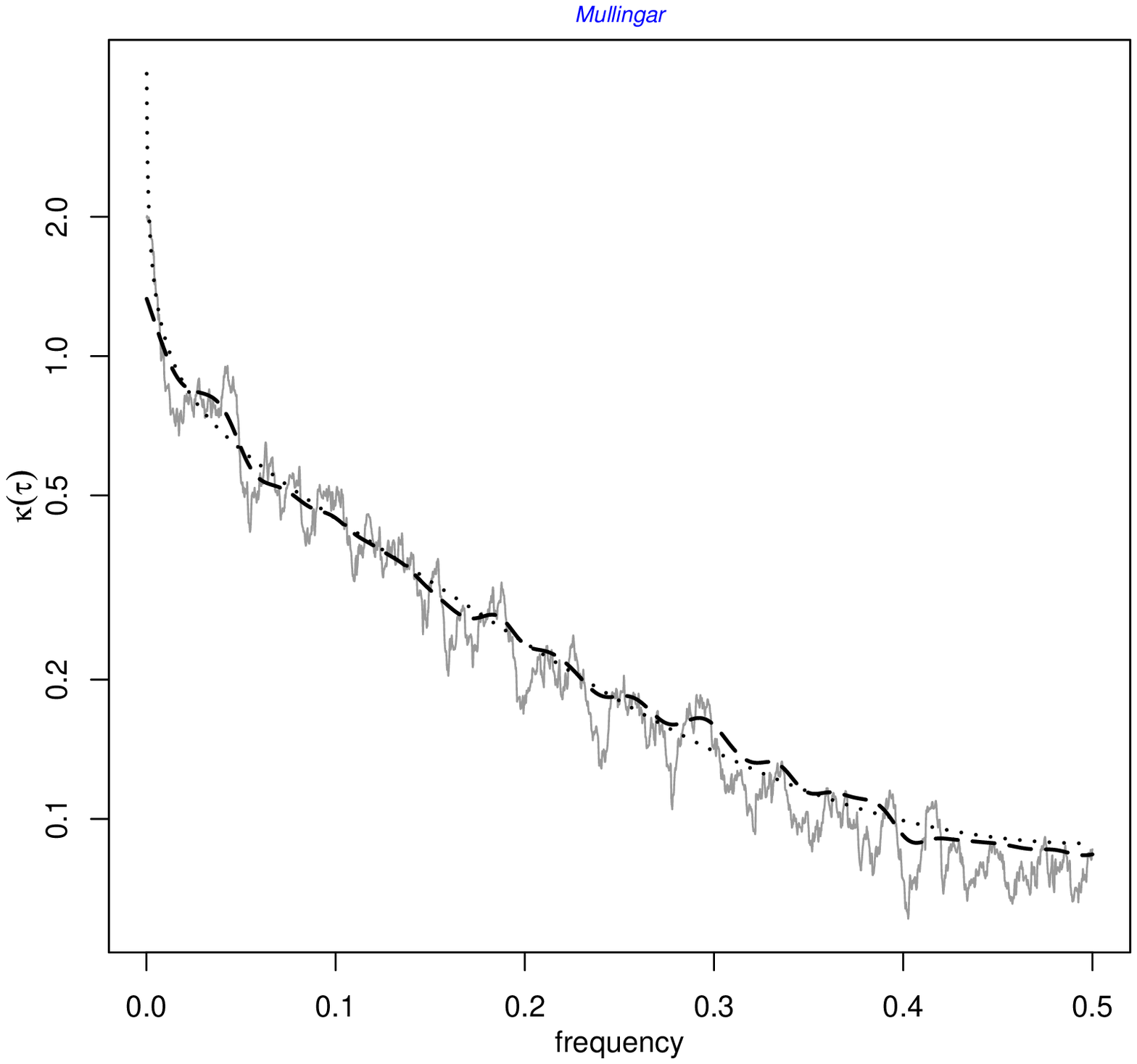}
\includegraphics[width=.325\columnwidth]{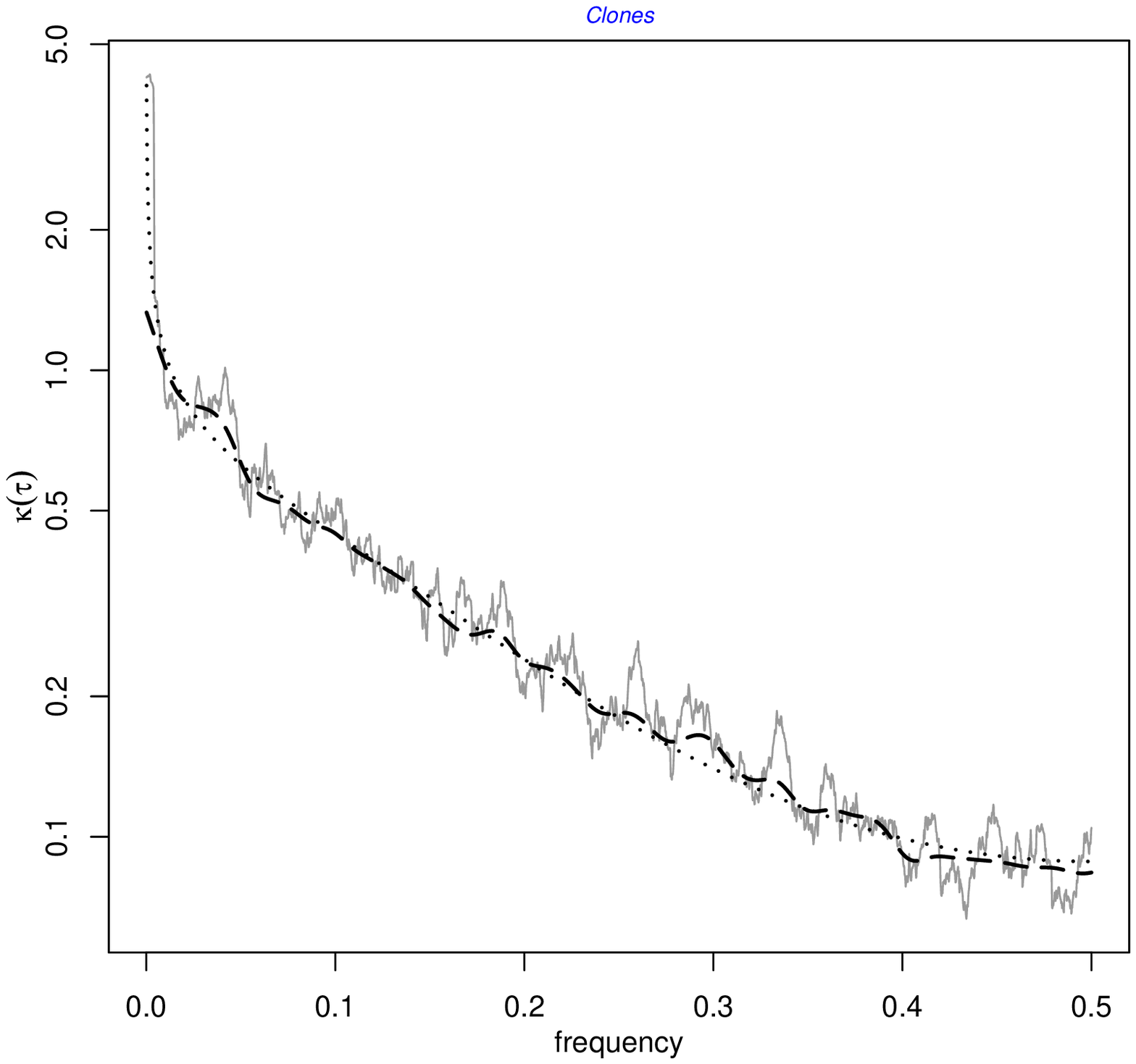}
\includegraphics[width=.325\columnwidth]{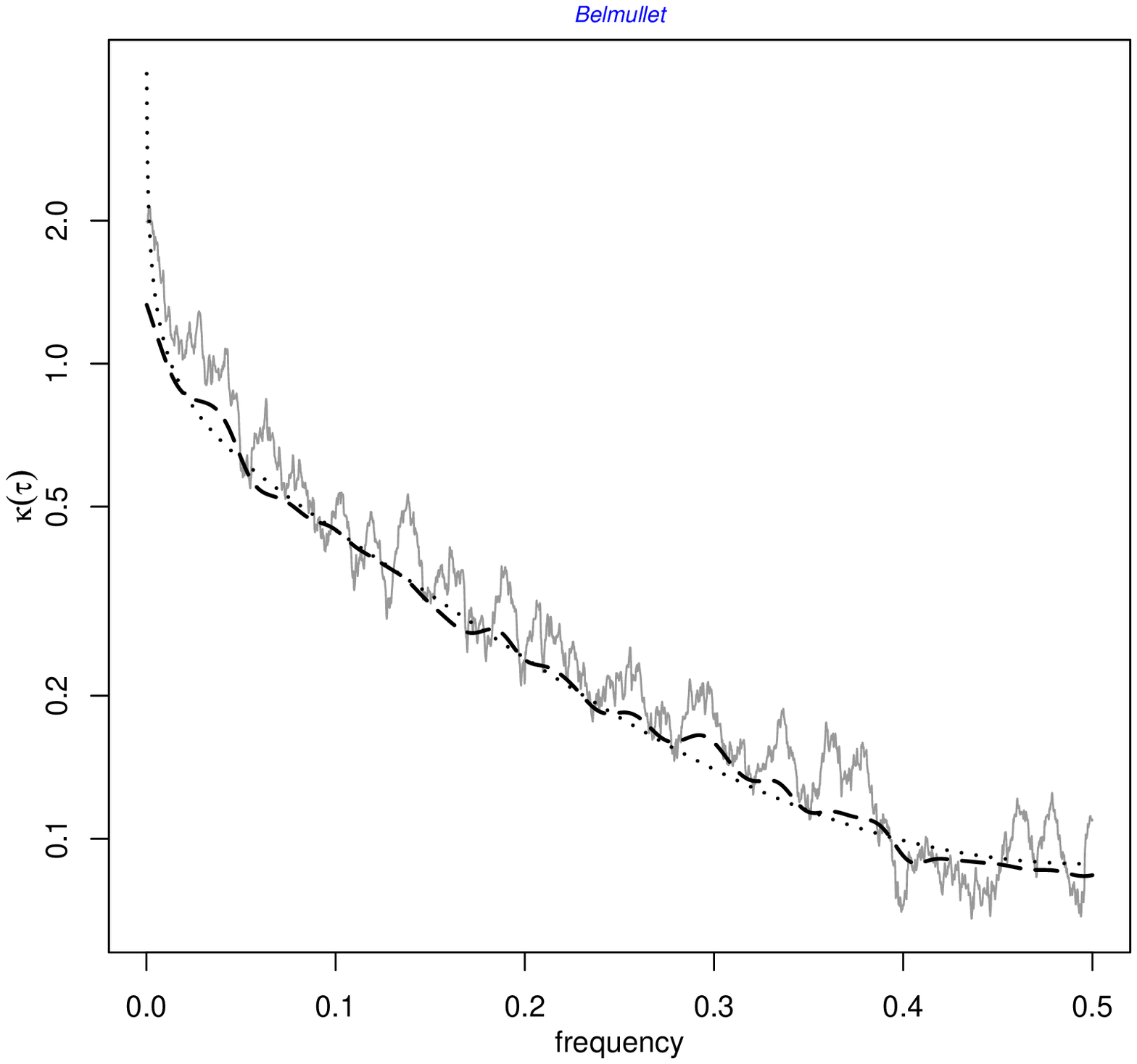}
\includegraphics[width=.325\columnwidth]{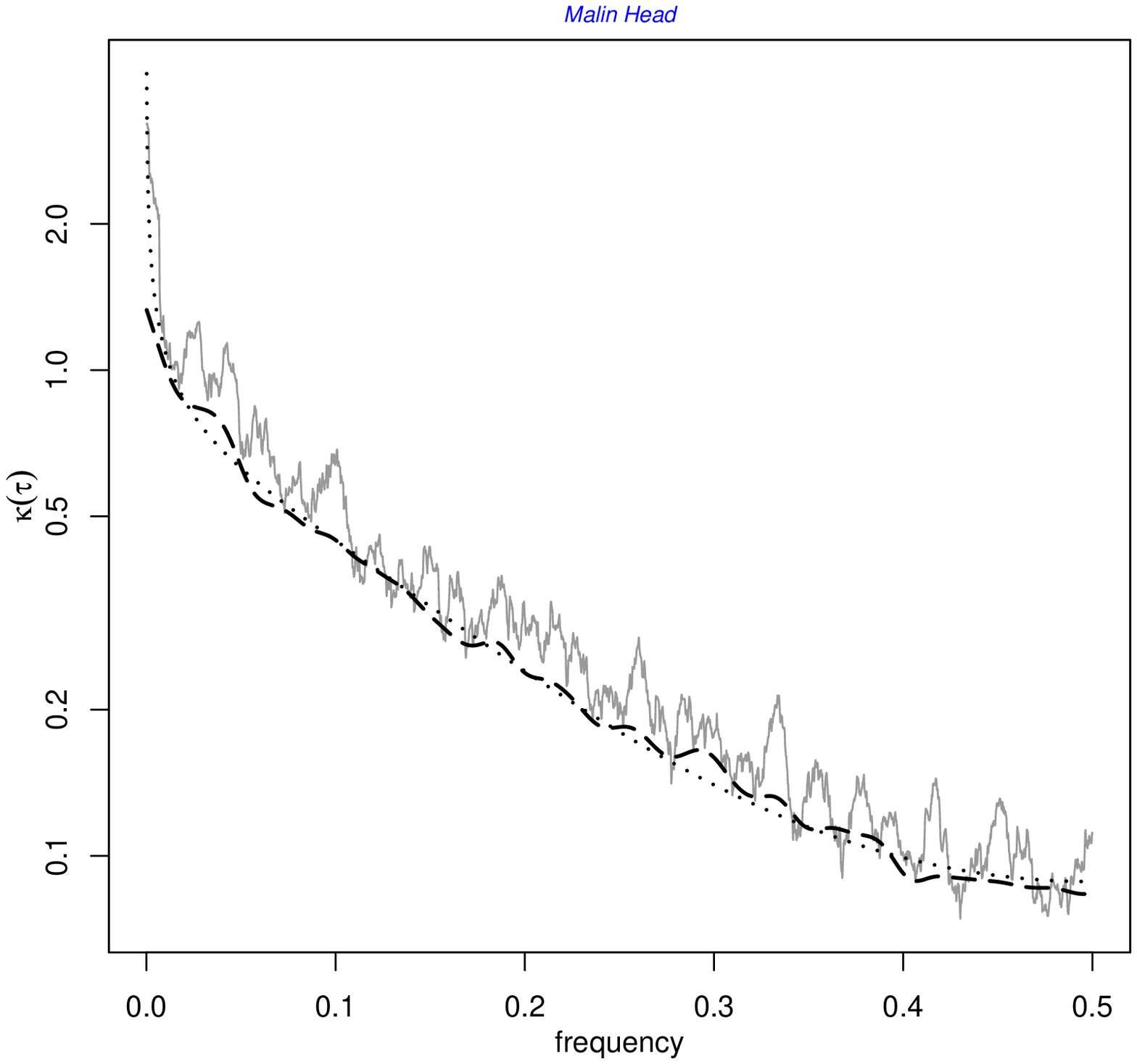}
  \caption{Smoothed empirical $\tilde{\tilde{k}}_i(\tau)$ (gray curves), parametric estimate
  $\hat{k}_{{par}}(\tau)$ (dotted
curves),   and nonparametric Nadaraya-Watson estimate
$\hat{k}_{{np}}(\tau)$ (long-dash curves) versus frequency for
the 11 individual  stations and their average marginal spectral,
$\tilde{\tilde{k}}(\tau)$ (first plot), by the standard program
tt{spec.pgram} in tt{R} with the span set to 5 for average
marginal spectra and  55 for individual marginal spectra.
}\label{spectrum}
\end{figure}

\begin{figure}
    \centering
\includegraphics[width=.45\columnwidth]{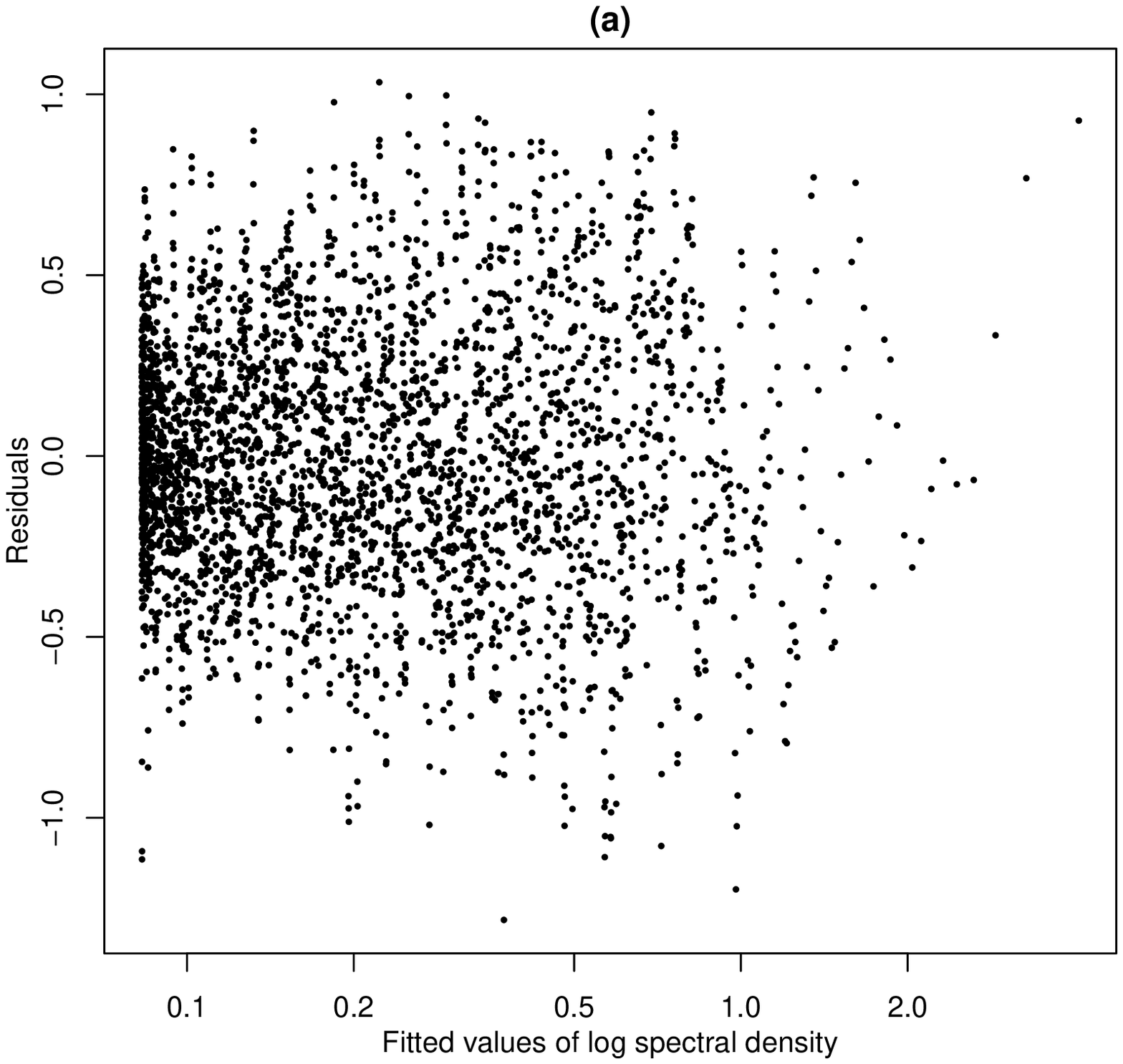}\hspace{.5cm}
\includegraphics[width=.45\columnwidth]{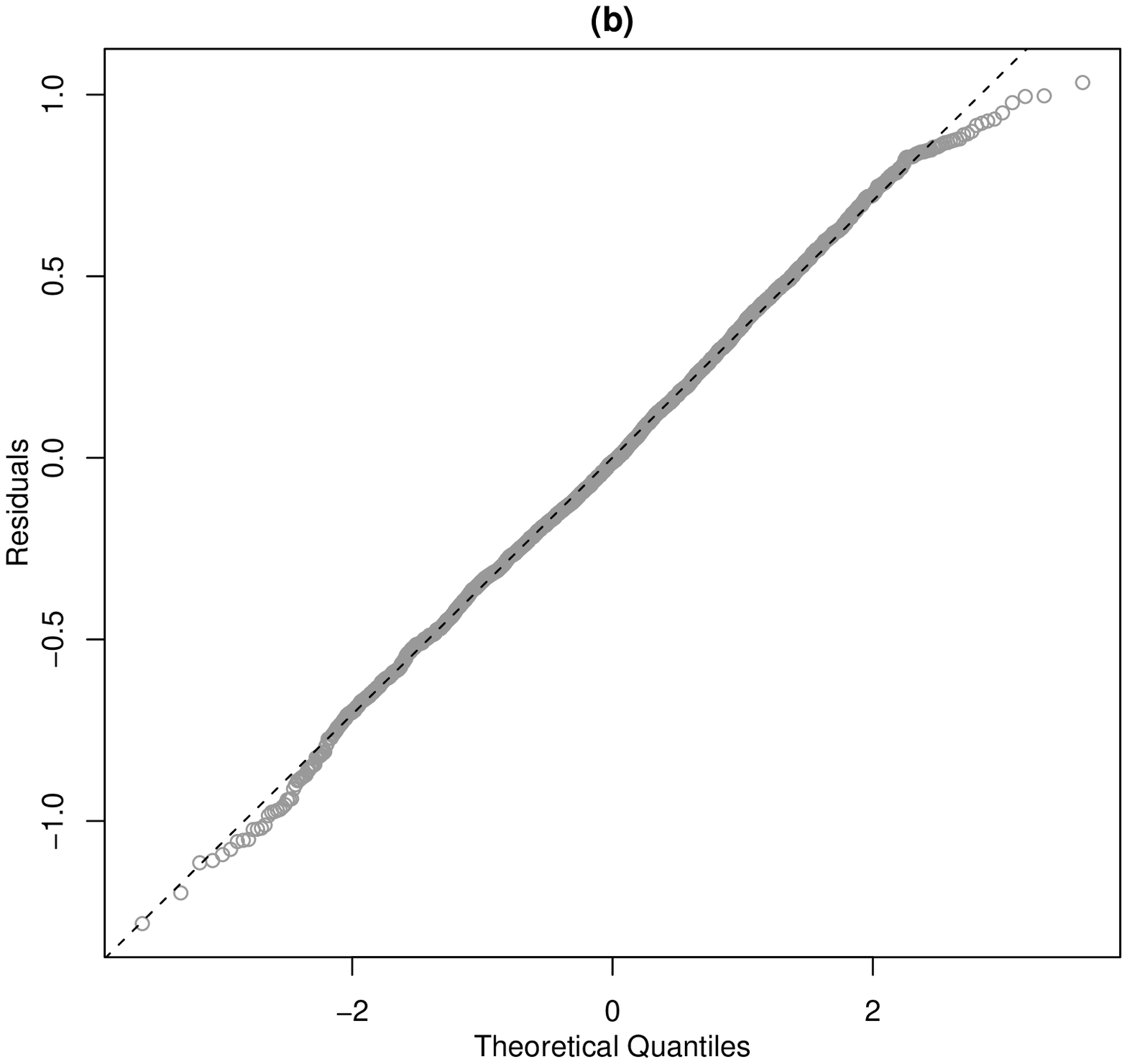}
  \caption{ (a): Residuals versus logarithm of fitted values of spectral density,
  $\log\hat{k}_{{par}}(\tau)$.  (b) normal Q-Q plot of residuals. }\label{spectq}
\end{figure}

\begin{figure}
    \centering
    \includegraphics[width=.45\columnwidth]{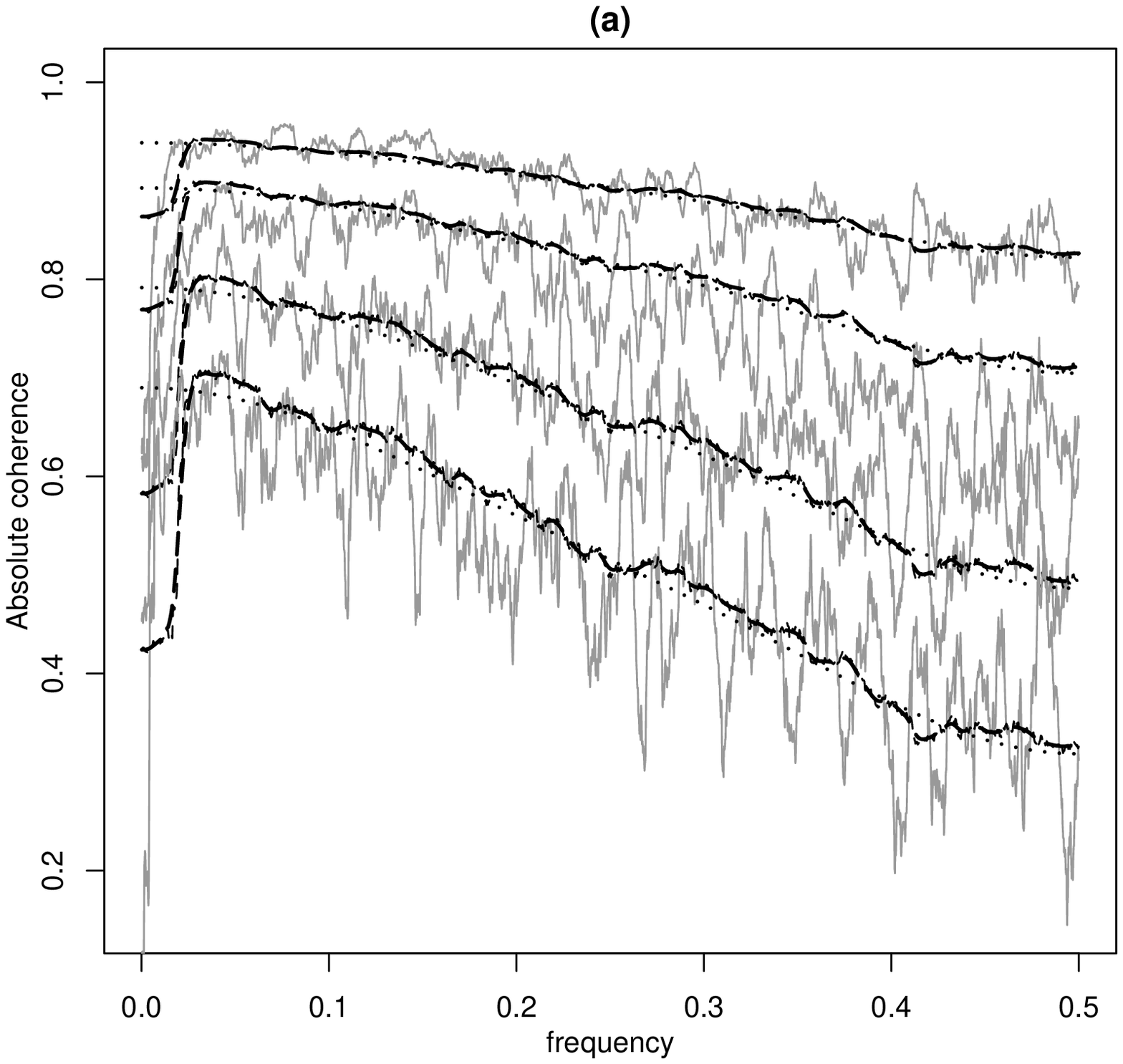}\hspace{.5cm}
    \includegraphics[width=.45\columnwidth]{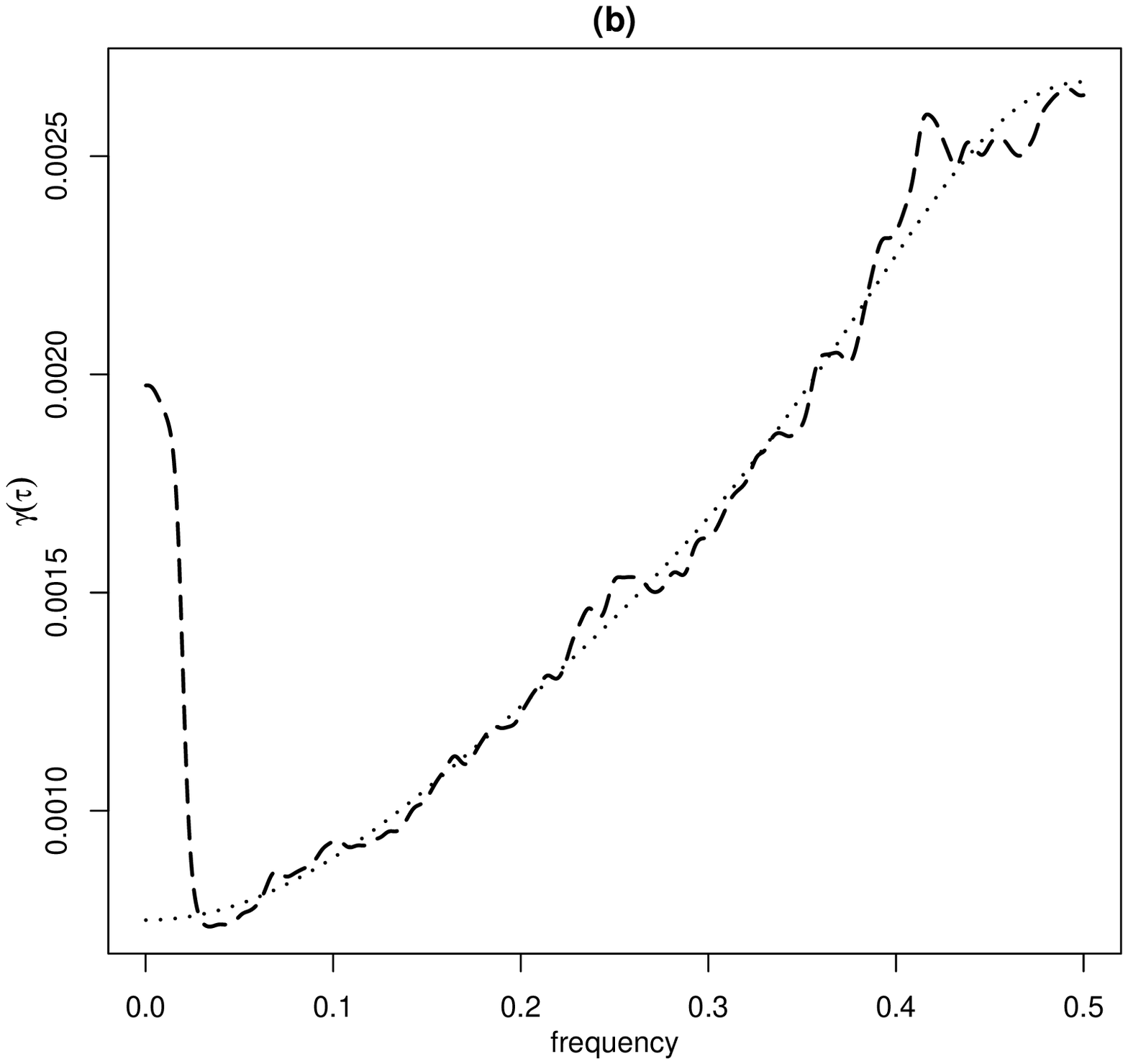}
  \caption{  (a): plot of empirical and estimated absolute coherence for different stations; from top:
  Birr-Mullingar: $|h|=60.68$ km,
  Birr-Dublin: $|h|=115.40$ km, Birr-Malin Head: $|h|=256.40$ km and  Valentia-Malin
  Head:  $|h|=427.34$
  km.  (b): plot of estimated   $\gamma(\tau)$. Empirical estimate (gray curve), Nadaraya-Watson
  estimate (long-dash curve)
 and trigonometric regression estimation (dotted curve). Empirical estimate calculated by the standard
 program
tt{spec.pgram} in tt{R} with the span set to 255.
   }\label{steincoh1}
\end{figure}

We now fit the Stein's asymmetric model (\ref{steincov}), with unknown
parameters $k(\tau)$, $\gamma(\tau)$, $\theta(\tau)$, $v$ and $p$ to
the Irish wind data.  The initially smoothed version
$\tilde{\tilde{H}}(h,\tau)$ of the empirical covariance-spectral
function $\tilde{H}(h,\tau)$ has been constructed using the standard
program tt{spec.pgram} in tt{R} with the span chosen
subjectively. For estimation purposes as well as plotting  figures the common choice t{span =255} has been used, though other choices have also been used below for exploratory purposes.

 \emph{Estimation of $k(\tau)$}:
 First we estimate $k(\tau)$. A stationary process with
spectral density having a pole at zero frequency is called a
stationary process with long-range dependence
\citep[Ch.2]{Beran1994}. Figure \ref{spectrum} shows smoothed
empirical spectral densities for the 11 individual stations,
$\tilde{\tilde{k}}_i(\tau)$,  along with their overall spectral
density $\tilde{\tilde{k}}(\tau)$. Since the averaging operation
will tend to smooth out the noise in individual periodograms, we
have chosen a smaller smoothing parameter for the averaged spectrum
rather than for the individual periodograms. The plots of the
marginal spectral densities  show the spectral densities have
roughly the same form at all stations with an apparent pole at zero
frequency suggesting the existence of long-range dependence;
Therefore Stein's parametric fractional exponential model
(\ref{ktaumodel}) seems to be an appropriate model for the
long-range dependence spectral density $k(\tau)$.  The model selection
criterium   AIC suggests $K_1=3$ is appropriate choice in
(\ref{ktaumodel}).  For $K_1=3$, letting $Y=\left(\log
\tilde{\tilde{k}}(\tau)\right)$, $X=[1, -\log\sin(|\pi\tau|), \cos(2\pi\tau), \cos(4\pi\tau), \cos(6\pi\tau)]$, $\beta=(c_0, \beta,  c_1, c_2, c_3)'$ and  regressing $\log
\tilde{\tilde{k}}(\tau)$ on the right hand side of (\ref{ktaumodel})
by OLS yields estimates $\hat{\beta}=0.315\pm 0.115,
\hat{c}_0=-1.769\pm 0.092,\hat{c}_1=0.710\pm  0.132,
\hat{c}_2=0.022\pm  0.086, \hat{c}_3=0.033\pm 0.074$.
Here the
intervals are based on 2 standard errors fitted by the procedure
developed in the previous section taking into  account the possibly
correlated errors.

  The fitted parametric spectral density
$\hat{k}_{par}(\tau)$ and the fitted nonparametric
Nadaraya-Watson estimate $\hat{k}_{np}(\tau)$ are plotted in
Figure \ref{spectrum}. As can be seen in Figure \ref{spectq}, it
seems that the normality and homoscedasticity assumptions  on the residuals are
sufficiently satisfied to justify the parametric OLS estimation.

\emph{Estimation of coherence}:  Next  we  consider empirical
coherence plots for various pairs of sites. Figure
\ref{steincoh1}(a) shows a plot of the  smoothed absolute coherence
versus time frequency for a subset of  spatial lags including the
biggest and smallest spatial lags. Our investigation indicates that
the optimal smoothing parameters  of coherence do not differ
significantly from each other
 and so we use a common smoothing parameter
for different spatial lags.
 This plot shows
that  the coherence  decays exponentially with the decay parameter
depending on the spatial lag. Therefore we choose the parametric
power exponential function (\ref{parametriccoh}). The low coherence
at low frequencies seems to be due to the long-range dependence in
time.   We omitted the first 300 frequencies in our estimation
procedure.

\begin{figure}
    \centering
\includegraphics[width=.35\columnwidth]{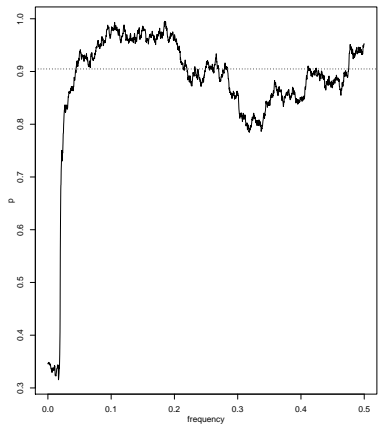}
 \caption{Least squares estimate of $p$   from separate slope model (\ref{newgamma1}) for different
 time frequencies. The dotted line is the estimated common slope}\label{pestimate}
\end{figure}

\emph{Estimation of $\gamma(\tau)$}: In the next step we estimate
$\gamma(\tau)$ and $p$. First in (\ref{newgamma1})  we assess the
validity of common slope model by fitting a regression model with a
separate slope for each $\tau$ and investigate how the fitted slopes
behave. Figure \ref{pestimate} indicates that the fitted value of
$p$ are nearly constant over the time frequency, so  a
parallel-lines regression is appropriate. Thus we use common slope
model (\ref{newgamma1}).

\begin{figure}
    \centering
\includegraphics[width=.45\columnwidth]{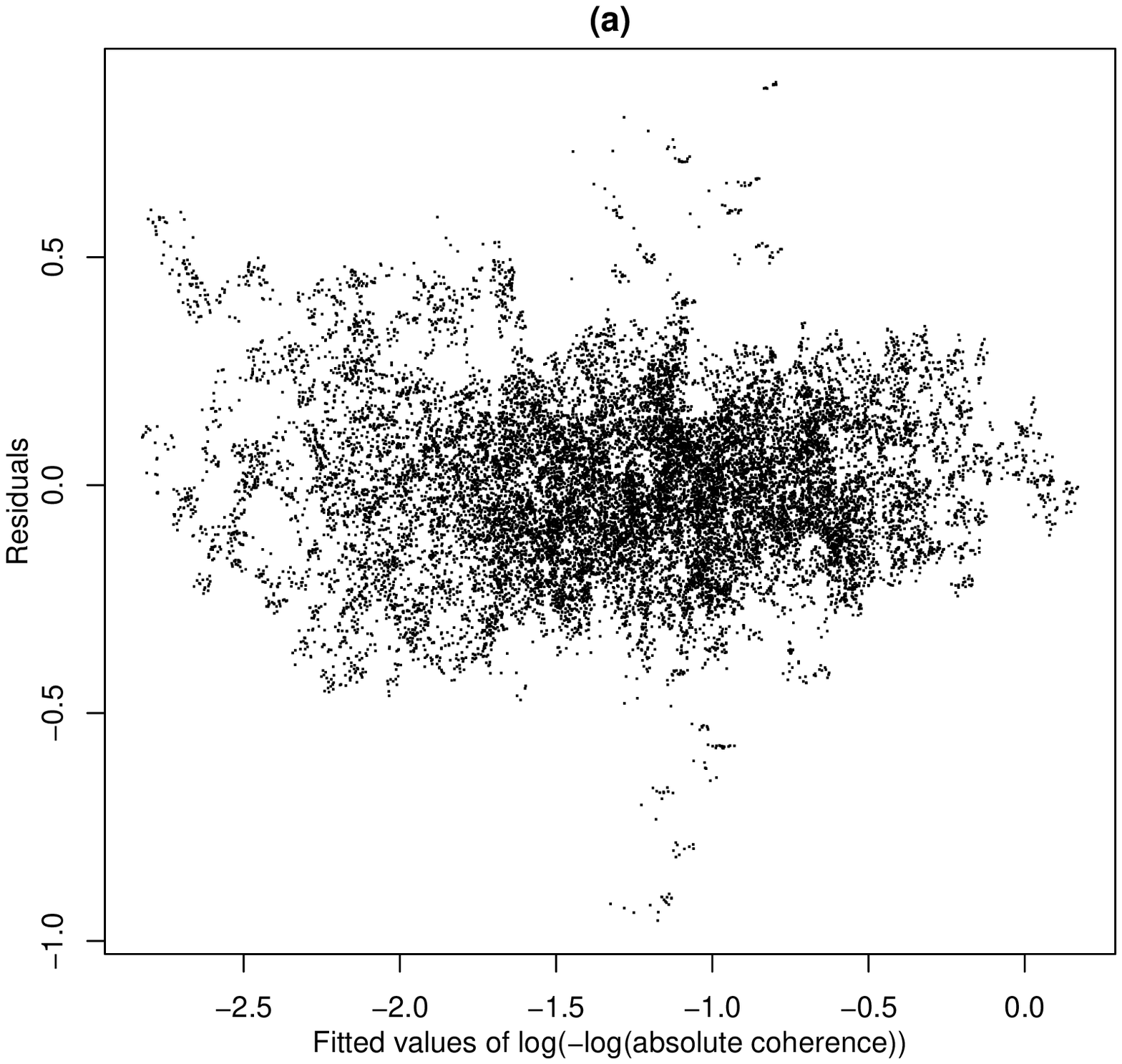}
\includegraphics[width=.45\columnwidth]{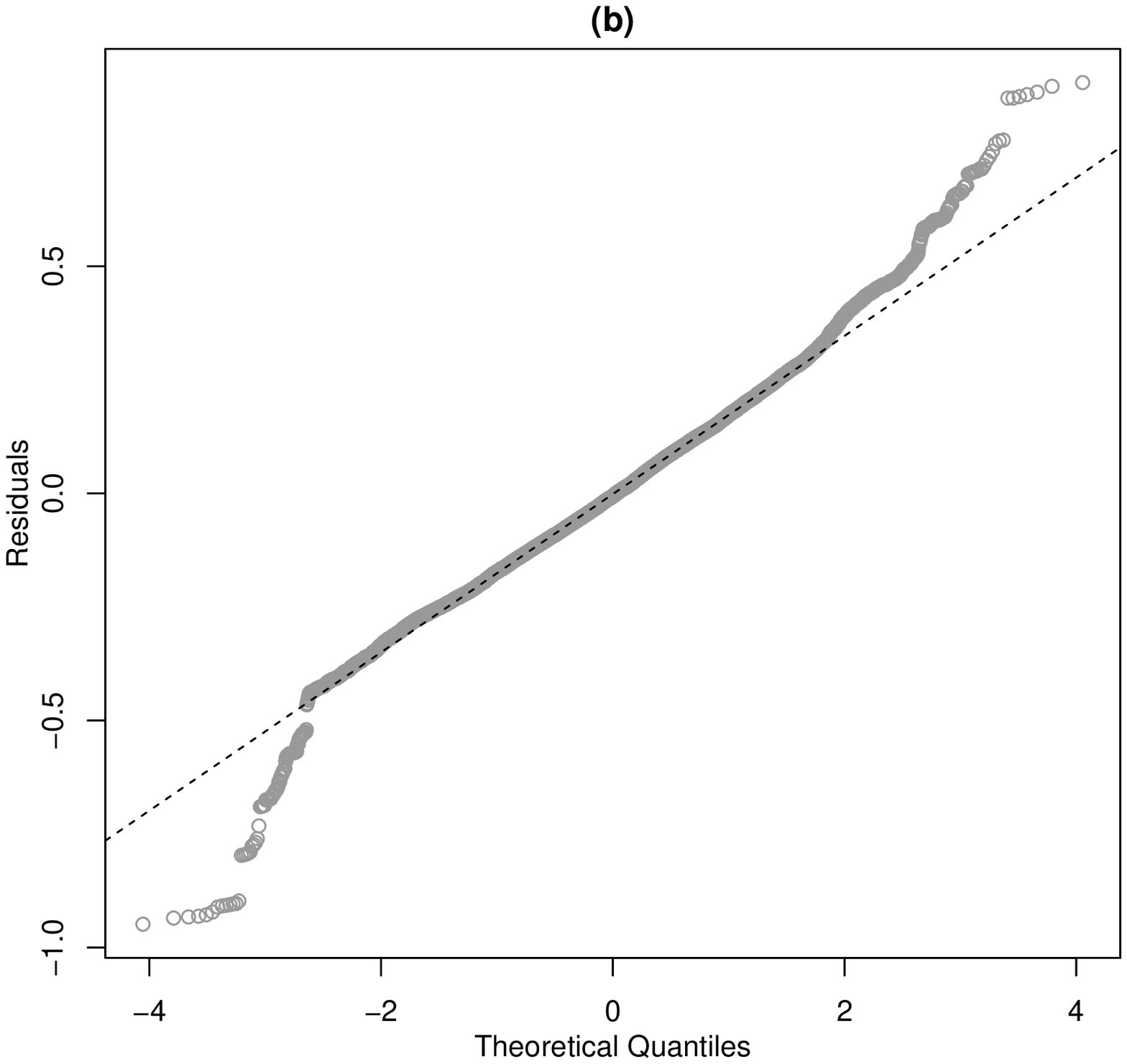}
 \caption{(a): plot of  residuals versus fitted  $\log(-\log( { absolute coherence}))$. (b):
 normal Q-Q plot of residuals. }\label{boxplots}
\end{figure}

The estimated common slope is
$\hat{p}=0.905\pm 0.005$. For $K_2=3$,  the estimated coefficients
are given by $\hat{a}_0=-6.551\pm 0.019,\ \hat{a}_1=-0.594\pm 0.028,\
\hat{a}_2=0.010\pm 0.027,\ \hat{a}_3=-0.042\pm 0.026$ which define
$\hat{\gamma}_{{par}}(\tau)$. The parametric and
Nadaraya-Watson estimates are presented in Figure
\ref{steincoh1}(b). By model (\ref{gammamodel}), estimates of  $p$
and $\gamma(\tau)$ define estimate of  the absolute coherence for
any spatial lag i.e. see Figure \ref{steincoh1}(a). The agreement
with the empirical absolute coherence looks reasonable. Figures
\ref{boxplots} depicts residuals for the empirical version of
(\ref{newgamma1}) and indicates that the homoscedasticity assumption
of the $\log(-\log( { absolute coherence}))$ transform is plausible.

\begin{figure}
    \centering
\includegraphics[width=.45\columnwidth]{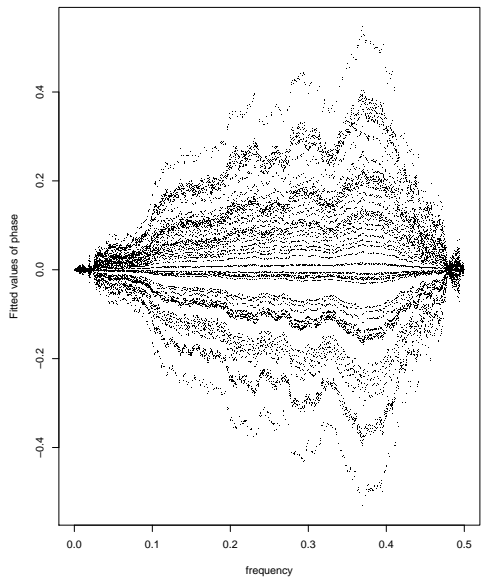}\hspace{.5cm}
 \caption{  plot of smoothed empirical phase for all  different pairs of stations.}\label{allgammaav}
\end{figure}

\begin{figure}
    \centering
\includegraphics[width=.45\columnwidth]{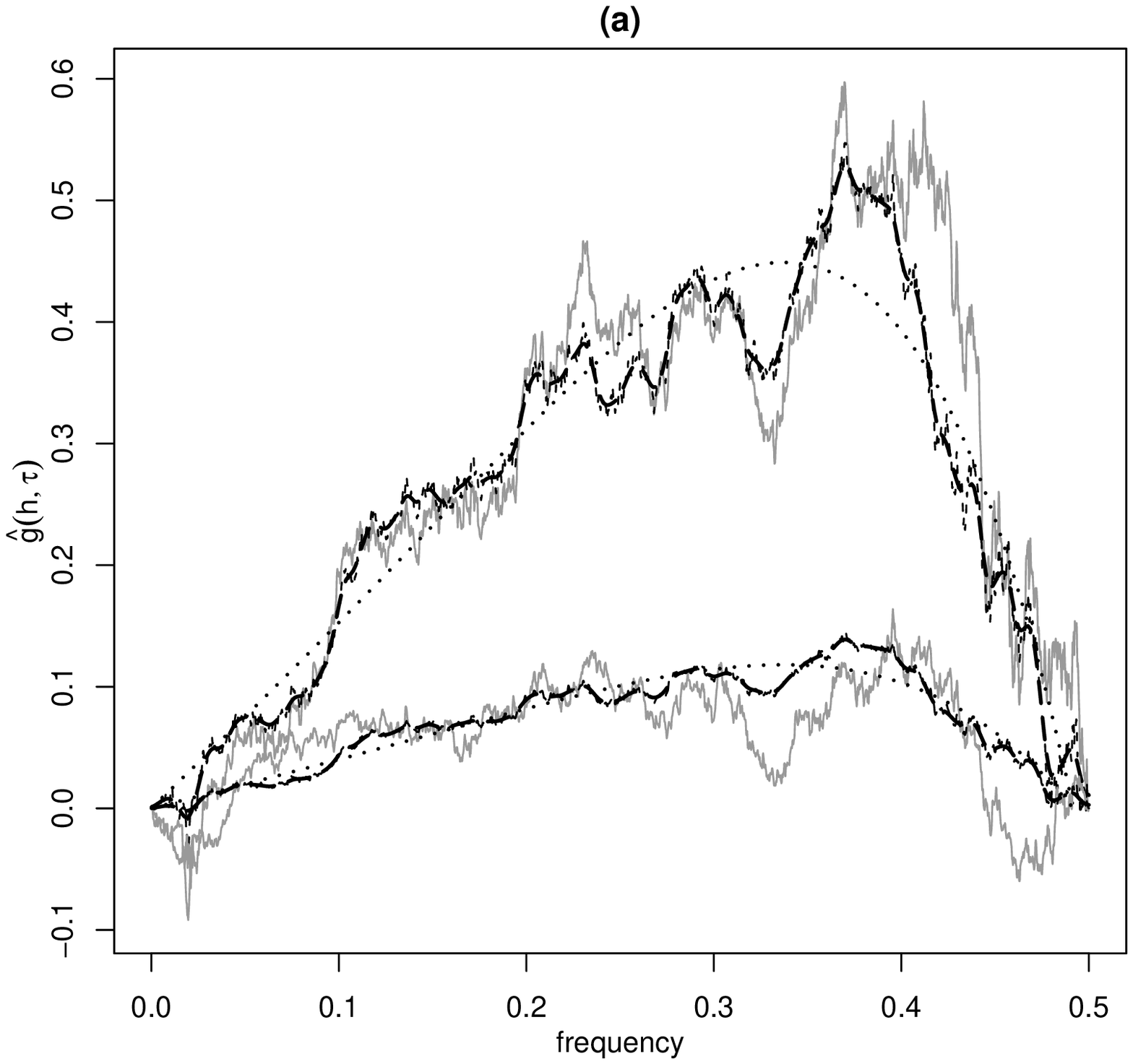}\hspace{.5cm}
\includegraphics[width=.45\columnwidth]{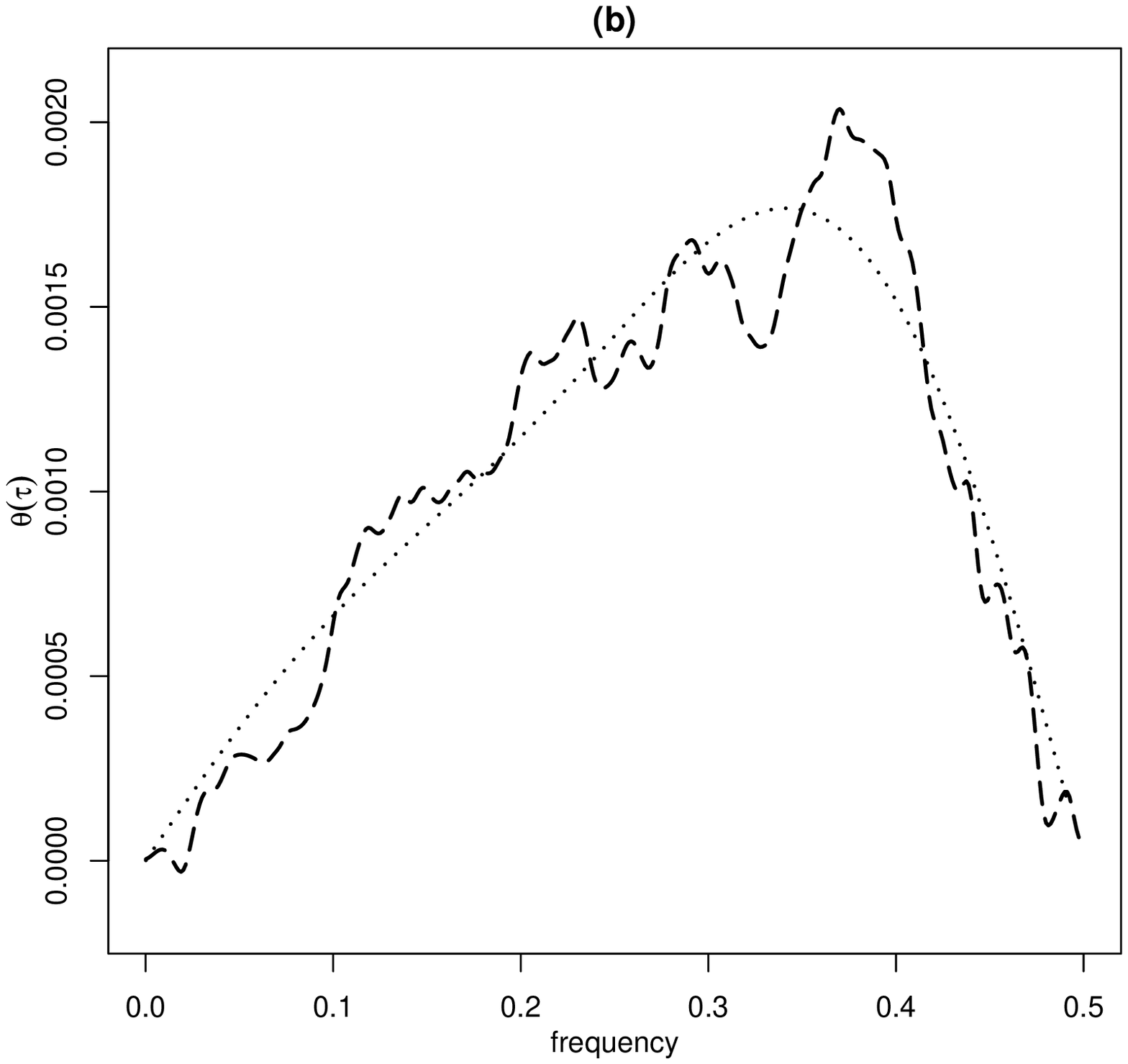}
 \caption{ (a): plot of empirical and estimated
phase for two different pairs of stations; top: Valentia-Dublin with
distance  316.99 km and bottom:  Clones-Dublin with distance 105.47
km. (b): plot of  estimated $\theta(\tau)$. Empirical estimate (gray
curve), Nadaraya-Watson estimate (long-dash curve) and trigonometric
regression estimation (dotted curve).}\label{gammaav}
\end{figure}

\emph{Estimation of $\theta(\tau)$}:

Figure \ref{allgammaav} shows a plot of the empirical phase
$\tilde{\tilde{g}}_R(h,\tau)$ versus time frequency $\tau$ for all
different spatial lags.  All the curves lie above the horizontal axis,
but to improve visibility, half of them have been plotted below the
axis.  For each curve, the phase seems to increase linearly in $\tau$
for small $\tau$, but is pulled back to 0 at $\tau = 0.5$ due to the
periodic boundary conditions.  Note that the maximum absolute value on
the vertical axis is well below $\pi=3.14$ so there is no distinction
between the angular variable $\tilde{\tilde{g}}(h,\tau)$ and its
real-valued extension $\tilde{\tilde{g}}_R(h,\tau)$ for this
dataset.  As illustrated in Figure \ref{gammaav}(a), the slope of each
curve depends on the spatial lag $h$.

\begin{figure}
    \centering
\includegraphics[width=.45\columnwidth]{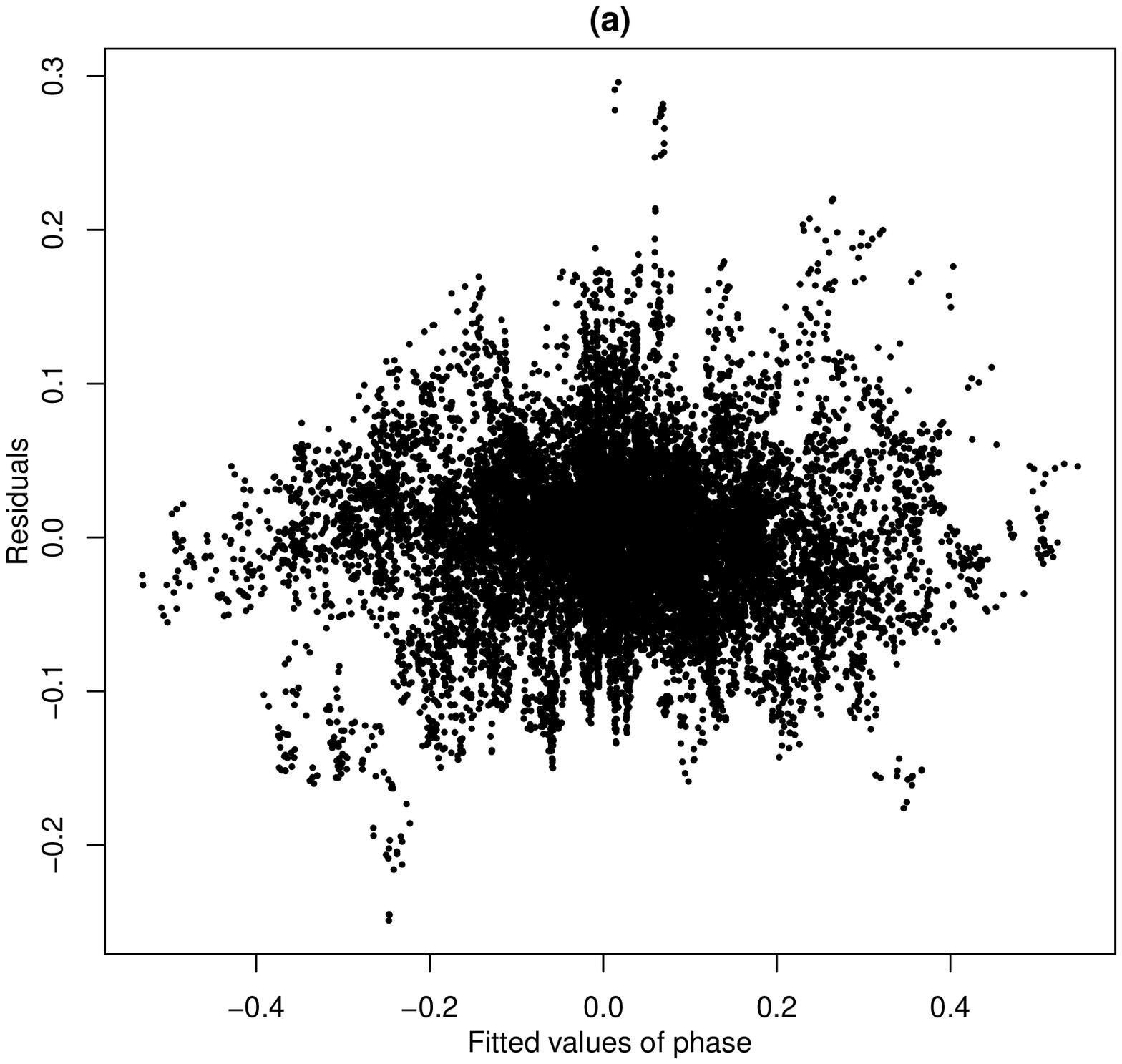}\hspace{.5cm}
\includegraphics[width=.45\columnwidth]{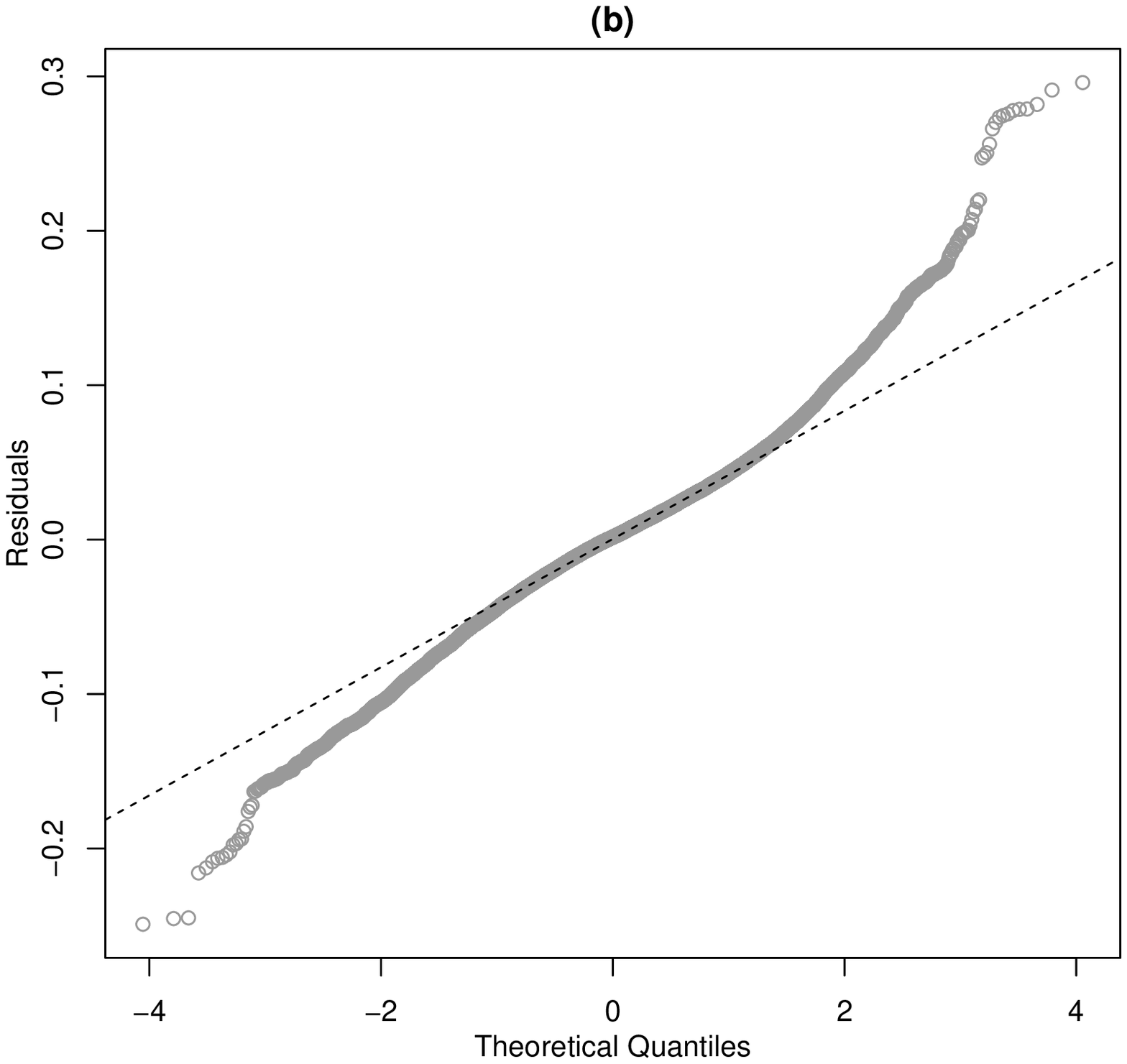}
  \caption{(a): Residuals versus fitted values of $\theta(\tau)$. (b): Normal Q-Q plot of residuals.
  }\label{phaseres}
\end{figure}

Using the estimation procedure in Section \ref{ch5sec6}, the optimally
estimated wind direction is given by $\hat{v}=(0.999 , 0.038)'$, i.e.
winds in Ireland are predominantly westerly.  Once $v$ has been
estimated, the initial estimate $\hat{\theta}_{{init}}(\tau)$ is
given by (\ref{ols}). Adopting a similar modelling strategy to that
used for $\gamma(\tau)$, we model the function $\theta(\tau)$ with a
trigonometric polynomial (\ref{thetamodel}).  After regressing
$\hat{\theta}_{{init}}(\tau)$ on $\tau$, the optimal order is
found to be $K_3=2$ by the AIC criterion, with estimated coefficients
$\hat{b}_1=0.00159\pm 0.05021, \ \hat{b}_2=-0.00045\pm  0.04022$
which define $\hat{\theta}_{{par}}(\tau)$. The result is
displayed in Figure \ref{gammaav}(b) along with the nonparametric
estimate. The parametric curve, the empirical curve, and the
Nadaraya-Watson curve all fit the phase reasonably well. Figure
\ref{gammaav}(a) shows a plot of the estimated phase versus time
frequency for two different spatial lags. Figure \ref{phaseres}, shows
that that the normality and homoscedasticity assumptions of residuals
are broadly satisfied for parametric estimation.

%


In this paper we give a deeper insight into the covariance-spectral modelling strategy and its properties. We proposed a simple transformation  on the covariance-spectral function to make it linear  in the unknown parameters  to use regression analysis. The method of estimation proposed here is more intuitive and  easier than Stein's method to apply for spatial-temporal data.
Effect  and  the amount of initial smoothing on the fitted function estimates and their standard errors are explored.  A very neat estimation for drift
 direction is proposed while Stein assumes it is known. The phase winding is  another important issue which is explored clearly in this paper.

 Stein constructed various plot to assess the goodness of fit of the model, we use similar plots to estimates the parameters.  Further, our method can be seen as more graphical, enabling visual judgments to be made about the estimated functions as can be seen e.g. for estimated $\gamma(\tau)$, $\theta(\tau)$ and $p$ as well as residual plots for goodness of fit assessment.
In general the behaviour of  our estimates matches  Stein's estimates very well. Since Stein fits data on sphere and uses different amount of smoothing,  numerical  comparing our estimates with that obtained by Stein is not appropriate.

\bibliography{references}
\bibliographystyle{apalike}
\end{document}